\documentclass[fleqn,usenatbib]{mnras}

\usepackage{newtxtext,newtxmath}
\usepackage[T1]{fontenc}
\usepackage{ae,aecompl}

\usepackage{graphicx}	% Including figure files
\usepackage{amsmath}	% Advanced maths commands

\usepackage{tabularx}   %Needed for vertical lines in table, without this c|c|c| not working
\usepackage{longtable}  %
\usepackage{adjustbox}  %for rotated table etc
\usepackage{pdflscape}
\usepackage[graphicx]{realboxes}
\usepackage{rotating}
\usepackage[normalem]{ulem} %ulem package for strikethrough text

\graphicspath{{./}{figures/}}

\title[Evolution of Li in red giants]{Lithium in red giants: the roles of the He-core flash and the luminosity bump}

\author[Deepak \& Lambert, D. L.]
{Deepak$^{1,2}$\thanks{E-mail: deepak@iiap.res.in, deepak4astro@gmail.com}
and David L. Lambert$^{3}$\thanks{E-mail: dll@astro.as.utexas.edu}\\
$^1$Indian Institute of Astrophysics, Bangalore - 560034, India\\
$^2$Pondicherry University, R. V. Nagara, Kalapet, Puducherry - 605014, India\\
$^3$W.J. McDonald Observatory and Department of Astronomy, The University of Texas at Austin, Austin, TX 78712, USA}
\date{Accepted 9$^{th}$ July 2021. Received 9$^{th}$ July 2021; in original form 31$^{st}$ March 2021}

\pubyear{2021}

\begin{document}
\label{firstpage}
\pagerange{\pageref{firstpage}--\pageref{lastpage}}
\maketitle

\begin{abstract}
Lithium abundances for red giants in the GALAH DR3 survey are studied. The rare examples of Li-enriched stars with abundances {\it A}(Li) $\geq 1.5$ are confirmed to be He-core burning stars belonging to or evolved from the red clump with similar masses and metallicity: $M\simeq 1.1\pm0.2{\rm M_\odot}$ and [Fe/H] $\simeq-0.3\pm0.3$. Li enrichment over the Li abundance present in a star's predecessor at the tip of the red giant branch likely occurs in all these red clump stars. Examination of the elemental abundances (C to Eu) in the GALAH catalogue shows no anomalous abundances in red clump giants and, in particular, no dependence on the Li abundance, which ranges over at least five dex. Lithium synthesis is attributed to the He-core flash occurring in stars at the tip of the red giant branch. Models from the {\it Modules for Experiments in Stellar Astrophysics} ({\tt MESA}) match the observed evolution of these stars along the red giant branch and to the red clump but only at the low effective temperature end of the observed spread of red clump giants. Run of Li abundance on the red giant branch is fairly well reproduced by {\tt MESA} models. A speculation is presented that the series of He-core flashes not only leads to $^7$Li synthesis from a star's internal reservoir of $^3$He but also may lead to internal restructuring leading to the observed effective temperature spread of red clump stars at about a constant luminosity. Giants exhibiting marked Li enrichments are not found at other evolutionary phases and, in particular, not directly associated with the luminosity bump on the red giant branch for which the Li abundance increase does not exceed 0.3 dex.

\end{abstract}


%%%%%%%%%%%%%%%%%%%%%%%%%%%%%%%%%%%%%%%%%%%%%%%%%%

%%%%%%%%%%%%%%%%% BODY OF PAPER %%%%%%%%%%%%%%%%%%

\section{Introduction} \label{sec:introduction}

Lithium's presence in low mass red giants has occasioned considerable interest across the evolutionary phases from subgiant, the red giant branch, the red clump He-core burning giant to the asymptotic giant branch and then to the post-asymptotic giant branch followed by ejection of the envelope with the stellar residual present as a white dwarf. Lithium present at low abundance in a giant's main sequence progenitor is depleted in the giant because the deep convective envelope drastically dilutes the Li abundance. This major episode of dilution occurs as the giant evolves to the base of the first red giant branch (RGB). This stage is known as the first dredge-up (FDU). Continuing depletion of Li occurs in the lowest mass RGB giants beyond the completion of the FDU. Given that the effect of the FDU in low mass giants is to lower the envelope's Li abundance by approximately a factor of 100 (more precise estimates are provided below), a red giant post-FDU of about solar metallicity is expected to have a Li abundance of {\it A}(Li) $\sim 1.5$.\footnote{Elemental abundances are given on the traditional scale: {\it A}(X) = $\log$[{\it N}(X)/{\it N}(H)]+ 12 where {\it N}(X) is the number density of element X.} A Li abundance of about this value represents the anticipated {\it upper} limit for a giant's Li abundance because Li may have experienced depletion, dilution or diffusion in a main sequence star's outer envelope well prior to the FDU and the giant's convective envelope may post-FDU destroy additional Li. Thus, a Li-rich giant has been widely known as one with an abundance {\it A}(Li)$\geq 1.5$.

Discovery of Li-rich K giants was initiated by \cite{WallersteinSneden1982}'s finding of a strong 6707 \AA\ Li\,{\sc i} line in the post-FDU K III giant HD 112127 with the abundance {\it A}(Li) $\simeq 3.2$. Discoveries of additional Li-rich K giants across a range of Li abundances trickled out after 1982, but, recently, discoveries have been provided at an accelerated pace thanks in large part to surveys such as the {\it Gaia}-ESO, GALAH and LAMOST surveys \citep{CaseyRuchti2016MNRAS.461.3336C,DeepakReddy2019MNRAS.484.2000D,GaoShiYan2019ApJS..245...33G,MartellJeffrey2021MNRAS.505.5340M}.
Li-rich giants are rare, representing a per cent or so of K giants in the Galactic disk \citep{DeepakReddy2019MNRAS.484.2000D}.
At present, the most Li-rich giant known may be TYC 429-2097-1 with {\it A}(Li) = 4.5 \citep{YanShiZhou2018NatAs...2..790Y}.

The discovery of HD 112127 and others led to speculations about Li production in these red giants. Almost all proposed explanations involve Li production tied to a giant's interior layer of $^3$He. This He isotope at low abundance is present in a star from its birth with additional amounts produced in low mass stars by initial steps of the {\it pp}-chain in main sequence H-burning. Li production from $^3$He is straightforward for a nuclear astrophysicist: the reaction $^3$He($\alpha,\gamma$)$^7$Be is followed by electron capture $^7$Be(e$^-$,$\nu_e$)$^7$Li. For the stellar astrophysicist, this link requires coupling the hot reaction site where $^7$Li is created but also destroyed by protons to the cool stellar atmosphere where the Li may survive and is observed. This necessary coupling acknowledges too that the intermediary $^7$Be has a half-life of just 53 days. The giant's stellar convective envelope is required to bring the fresh ($^7$Be and) $^7$Li to the atmosphere with, of course, also convection returning $^7$Li to the interior for destruction and replenishment. This coupling of nuclear and stellar astrophysics first recognized by \cite{Cameron1955ApJ...121..144C} is now widely known as the Cameron-Fowler mechanism.

\cite{CameronFowler1971} suggested this mechanism involving  He-burning shell flashes to account for the rare carbon stars like WZ Cas \citep{McKellar1940PASP...52..407M} and WX Cyg \citep{Sanford1950ApJ...111..262S} with very strong 6707 \AA\ lines. In modern parlance, these carbon stars are examples of luminous AGB stars experiencing helium shell burning episodes and the third dredge-up. Lithium, carbon and $s$-process enrichment are predicted by contemporary modelling and observed in luminous AGB  giants \citep{SmithLambert1989}. However, for K giants like HD 112127, which are not luminous AGB stars,  two evolutionary phases have received the most attention with respect to Li enrichment: giants at the so-called luminosity bump (LB) on the RGB and more luminous low mass giants at the RGB's tip, which experience a He-core flash before evolving to the red clump (horizontal branch) as a He-core burning giant.

Suspicions about the association of the He-core flash with Li synthesis were aired in the first survey for Li-rich giants by \cite{BrownSnedenLambert1989ApJS...71..293B} and raised subsequently by others.
For example, \cite{DeepakReddy2019MNRAS.484.2000D} found that most of the Li-rich giants in the GALAH DR2 survey belong to the red clump phase.
%and then in the subsequent study \citep{DeepakLambertReddy2020MNRAS.494.1348D} found that the Li-rich giants are chemically and rotationally similar to the Li-normal giant, suggesting that the origin of Li enhancement in giants may lie at the RGB tip during the He-core flash rather than by external source of merging of sub-stellar objects or during luminosity bump evolution.}
These suspicions were wonderfully raised to a near-certainty by \cite{KumarReddy2020NatAs...4.1059K} whose analysis of Li abundances of giants in the GALAH DR2 survey led to the bold suggestion that {\it only} He-core burning giants were Li-rich and, thus, that Li synthesis occurs at or immediately following the He-core flash.
They asserted that Li enrichment of He-core burning giants is `ubiquitous' and confined to these giants, that is -- all low mass giants experiencing a He-core flash synthesize Li at a level from very mild to severe to contaminate the RC giant which follows the flash. Subsequent studies of asteroseismology of He-core burning giants with A(Li) determinations suggest that Li abundances following the He-core flash may decline with age \citep{DeepakLambert2021MNRAS.505..642.,SinghReddy2021ApJ...913L...4S}.
The common assumption is that the Li results from the Cameron-Fowler mechanism.  \cite{KumarReddy2020NatAs...4.1059K} did not detail the mechanism's application to the He-core flash. A recent theoretical exploration \citep{Schwab2020ApJ...901L..18S} is discussed below.

Adaptation of the $^3$He reservoir to Li production has also been connected to RGB giants experiencing the LB. This `bump' on the RGB occurs at the point when the H-burning shell moves outward across an H-discontinuity. The evolution of the star reverses course, and it moves slowly and slightly down the RGB before reversing again and continuing to evolve up the RGB. This manoeuvre introduces a clumping of stars on the RGB. At the time of this LB, operation of the Cameron-Fowler mechanism occurs through a thermohaline instability expected on theoretical grounds \citep{EggletonDearborn2008ApJ...677..581E,LagardeCharbonnel2011A&A...536A..28L} and suggested as a possible way to account for Li enriched giants \citep[e.g.,][]{MartellJeffrey2021MNRAS.505.5340M}. But definitive observational identification of Li enriched LB giants remains open.\footnote{Thermohaline instability at the LB has been suggested also as a way to change a giant's surface $^{12}$C/$^{13}$C ratio.}

In this paper, we concentrate on the exploitation of the recent GALAH survey of stellar abundances, with the focus here on the Li abundances of low-mass red giants. Our primary goal is to set Li production limits by the He-core flash as a function of stellar mass and initial metallicity. Our secondary goal is to search for Li enrichment by LB giants. Our observational inquiry based on the latest data release DR3 from the GALAH survey is supported by predictions of Li abundances from calculations performed using the Modules for Experiments in Stellar Astrophysics \citep[{\tt MESA};][]{PaxtonBildstenMESAI_2011ApJS..192....3P,PaxtonCantielloMESAII_2013ApJS..208....4P,PaxtonMarchantMESAIII_2015ApJS..220...15P,PaxtonSchwabMESAIV_2018ApJS..234...34P,PaxtonSmolec2019ApJS..243...10P} suite of computer programmes.

\section{Data sample} \label{sec:sample}

Recent observational advances in understanding Li-rich giants have largely come from combining spectroscopic abundance surveys with astrometric data from the {\it Gaia} satellite \citep{GaiaCollaborationTheGaiaMission2016A&A...595A...1G}. Li (and other) elemental abundances from the GALAH DR2 survey have been used by \cite{DeepakReddy2019MNRAS.484.2000D,DeepakLambertReddy2020MNRAS.494.1348D,KumarReddy2020NatAs...4.1059K} with \cite{MartellJeffrey2021MNRAS.505.5340M} using the DR3 results. The LAMOST survey was considered by \cite{DeepakLambertReddy2020MNRAS.494.1348D} for  C, N and O (but not for Li) for stars in common with the GALAH DR2. LAMOST's Li abundances were explored by \cite{SinghReddyKumarAntia2019ApJ...878L..21S} for stars in the Kepler field of view and in detail by \cite{YanZhouZhang2021NatAs...5...86Y}.

Here, we consider the third data release (DR3) of the GALAH survey. Our analysis is devoted to low mass giants with a metallicity [Fe/H] representative of the Galactic disk, say $-1 \lesssim$ [Fe/H] $\lesssim +0.5.$ GALAH DR3 (and DR2) provides too few metal-poor halo giants to probe the matter of Li-richness among the halo population. As a secondary interest, we make a  foray into possible abundance anomalies introduced into RC giants by the He-core flash, which converts a giant at the RGB tip to a He-core burning giant on the RC.

\begin{figure}
\includegraphics[width=0.5\textwidth]{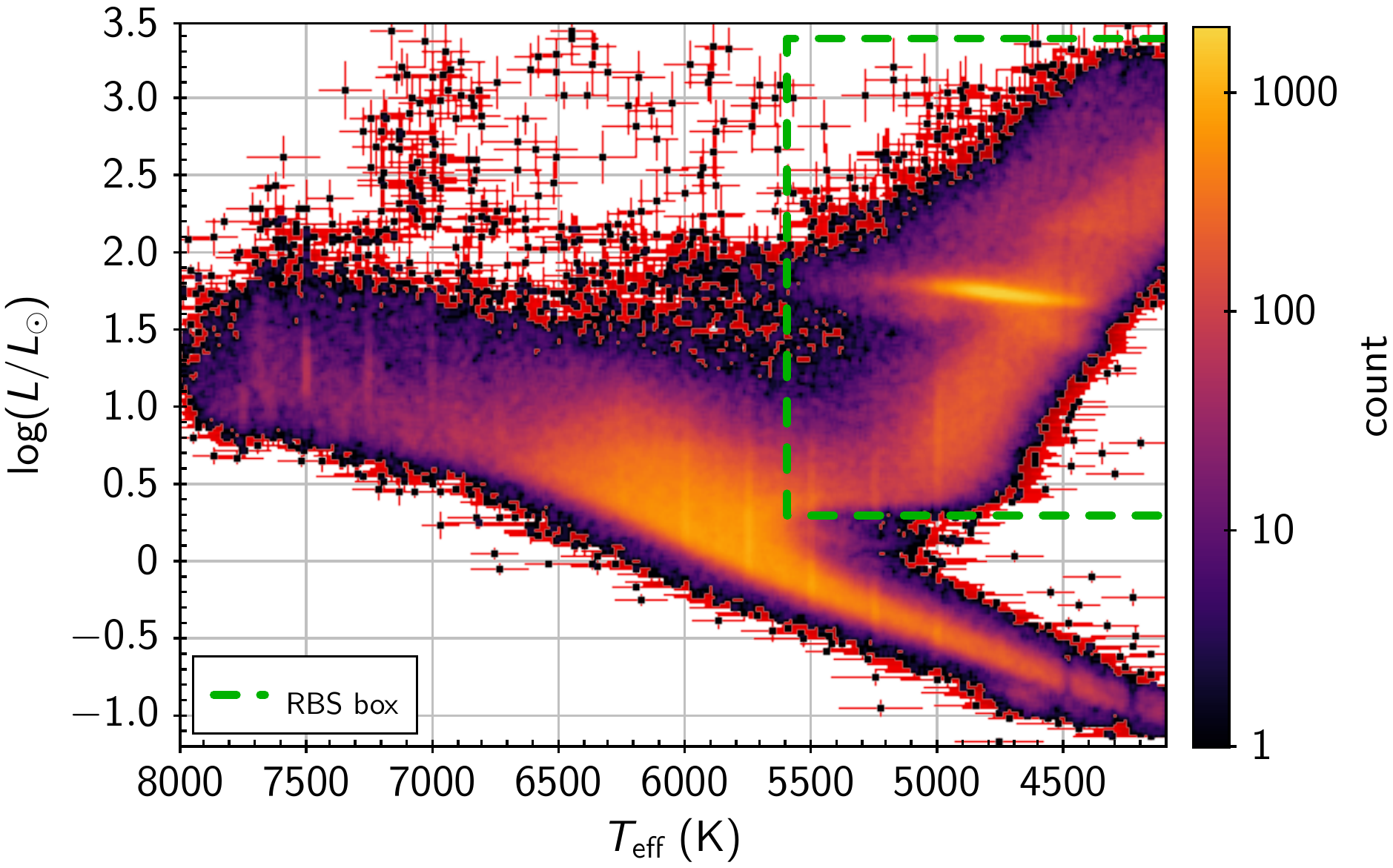}
\caption{Sample stars in a HR diagram also showing the selection of red giants (the RBS) within the green box. For all the stars, uncertainty in luminosity and temperature are also plotted in the background as red error bars.
\label{fig:1}}
\end{figure}

The Third Data Release of the Galactic Archaeology with the HERMES survey provides stellar parameters and elemental abundances derived from 678,423 spectra for 588,571 unique stars observed with the HERMES spectrograph at the Anglo-Australian Telescope between November 2013 and February 2019 \citep[see][for more detail on the data release]{GalahDR3_BuderSharma2021MNRAS.tmp.1259B}. It is also accompanied by several value-added-catalogues (VAC) providing various other intrinsic, kinematic and dynamical properties of the observed stars (hereafter, GALAH$+$ DR3).\footnote{The GALAH$+$ DR3 data is available at \url{https://cloud.datacentral.org.au/teamdata/GALAH/public/GALAH_DR3/}}
{\tt GALAH\_DR3\_main\_allspec\_v1.fits} is one of the two main catalogues of GALAH$+$ DR3 and includes stellar parameters and overall abundances of about 30 chemical species derived for each of the observed spectra. This catalogue also includes abundances derived for each individual line for all the elements. We joined this catalogue with the VAC {\tt GALAH\_DR3\_VAC\_ages\_v1.fits} which provides other stellar parameters like age, mass, distances, luminosity, etc., estimated using Bayesian Stellar Parameter Estimation code \citep[BSTEP,][]{SharmaStelloBuder2018MNRAS.473.2004S}. The two catalogues are joined using the recommended method of using the internal ID for each GALAH observation provided in the {\tt sobject\_id} column. To eliminate multiple entries of a star in the catalogue and to select the entry with the highest signal-to-noise, we restrict our sample to stars with {\tt flag\_repeat} = 0. We further restrict our sample to stars with reliable stellar parameters ({\tt flag\_sp} = 0) and metallicity ({\tt flag\_fe\_h} = 0) which lead to abandoning of stars with $T_{\rm eff} < 4100$ K.
We also discarded all the stars with uncertainty in temperature ($\sigma_{T\rm_{eff}}$) and luminosity ($\sigma_{\log(L/{\rm L_\odot})}$) of more than 200 K and 0.2, respectively. This resulted in a sample of about 425 thousand stars. Sample stars are shown in Figure \ref{fig:1} in the form of the HR diagram.  Note that the sample is not mass limited, and the large scatter in the figure can be attributed in part to the presence of relatively massive stars (with masses 2$-$5 M$_\odot$) in the sample.
For each sample star, uncertainties in luminosity and temperature are also shown in the background, however, due to crowding, they are visible only for the star on the boundaries. From this sample, we select giant stars with temperature from 4100 to 5600 K and luminosity $\log (L/{\rm L_\odot})$ from 0.3 to 3.4 dex. This selection of 159,841 stars is shown within the dotted green box in Figure \ref{fig:1} (hereafter, the RBS, which stands for the red-giant branch sample).

Except for \cite{MartellJeffrey2021MNRAS.505.5340M}, previous discussions of Li abundances in giants based on the GALAH survey drew on the DR2 catalogue. Of interest to the present study is the fact that our sample of RBS giants from the DR3 catalogue is approximately doubled over our previous sample from DR2 \citep{DeepakLambertReddy2020MNRAS.494.1348D}, abundance quality flagging is also improved, including the provision of upper limits and abundances are now provided for 31 rather than 21 elements previously. In the case of the 335 Li-rich giants found by \cite{DeepakReddy2019MNRAS.484.2000D} in the DR2 catalogue, Li abundances of common stars in DR2 and DR3 show only a slight systematic change: the least Li-rich are slightly less Li-rich, and the super Li-rich giants are slightly more Li-rich. Our assessment is that the differences between the data releases have not affected principal conclusions about Li in giants in the RBS collection studied in previous works.
The increase in the RBS sample offers the chance to search for new properties of Li-rich giants and to explore already covered properties in more detail.

\subsection{Elemental abundances}\label{sec:Abundances}

The GALAH$+$ DR3 survey provides the elemental abundance ratio [X/Fe]\footnote{For two elements, $X$ and $Y$, with number densities $N_X$ and $N_Y$, the elemental abundance ratio [$X/Y$] $\equiv$ $\log$($N_X/N_Y$)$\rm_{star}$ $-$ $\log$($N_X/N_Y$)$\rm_{Sun}$.}
for up to 30 chemical elements per star with X = Li, C, O, Na, Mg, Al, Si, K, Ca, Sc, Ti, V, Cr, Mn, Co, Ni, Cu, Zn, Rb, Sr, Y, Zr, Mo, Ru, Ba, La, Ce, Nd, Sm, and Eu. Abundances for most of the elements were considered in detail by \cite{DeepakLambertReddy2020MNRAS.494.1348D} using the DR2 catalogue. The global analysis of [X/Fe]  will not be reexamined here. Previously, the only detectable difference between [X/Fe] for Li-rich, Li super-rich and normal giants was that Li super-rich giants were slightly underabundant in [C/Fe] but see \cite{MartellJeffrey2021MNRAS.505.5340M}.

For Li, the DR3 catalogue provides the non-LTE correction for the Li abundance provided by the 6707 \AA\ resonance doublet. We convert the [Li/Fe] using the non-LTE Li and LTE Fe abundances into {\it A}(Li) using {\it A}(Li) = [Li/Fe] + [Fe/H] + {\it A}(Li)$_\odot$, where {\it A}(Li)$_\odot$ is the solar Li abundance for which we adopt {\it A}(Li)$_\odot$ = 1.05 $\pm$ 0.10 dex, which is the value used for the GALAH data \citep[see][for more details]{GalahDR3_BuderSharma2021MNRAS.tmp.1259B}.
The DR3 catalogue also provides uncertainties ($\rm \sigma_{[X/Fe]}$ available in corresponding {\tt e\_X\_fe} column) and a quality flag ({\tt flag\_X\_fe}) for each of abundance entry. {\tt flag\_X\_fe} can have any value from 0, 1 and 32, where flag 0 means provided abundance is reliable, 1 means provided abundance is an upper limit, and 32 means no reliable measurement reported. In this study, we will use {\tt flag\_X\_fe} = 0 for  elements except Li.
In the case of Li, {\tt flag\_Li\_fe} = 0 will be used to identify a Li-enriched star, while for the Li-normal stars, we will use all three flags to avoid discarding stars with a relatively lower Li-abundance from the sample.
Throughout the paper, we will specify if we are considering a sample of stars for which DR3 notes a detection of Li or considering the larger sample combining stars with a detection or an upper limit of the Li 6707 \AA\ line.

\section{Lithium-rich giants: overview}
\label{sec:analysis}

\begin{figure*}
\includegraphics[width=1\textwidth]{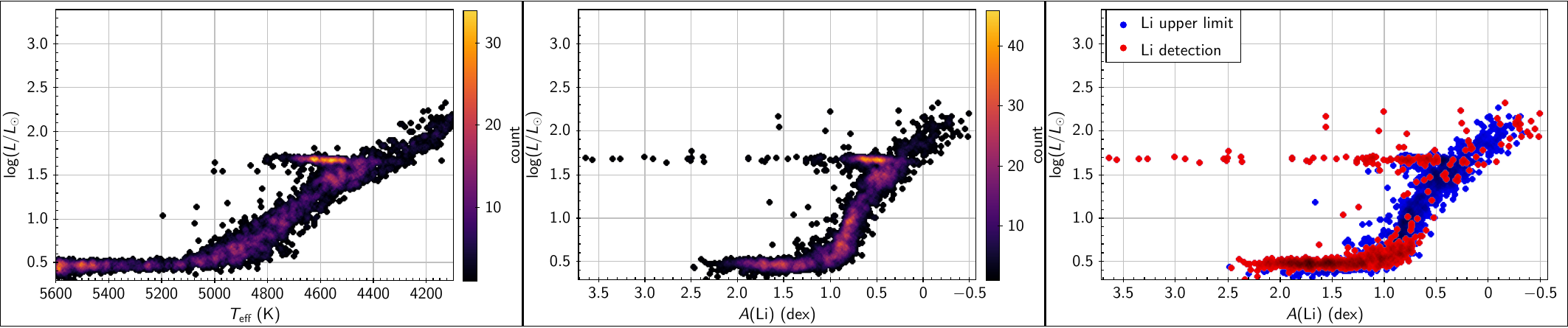}
\caption{HR diagram and {\it A}(Li) as a function of luminosity for the RBS giants with masses $M/{\rm M_\odot}$ = 1.10 $\pm$ 0.02 and metallicity [Fe/H] = 0.00 $\pm$ 0.10 dex.
The left and middle panels show giants with both Li detections and upper limits with light yellow colour indicating a higher density of stars. In the right-most panel, giants with Li detections are shown in red, while giants with Li upper limits are in blue.
\label{fig:HRD_and_LumALi_trends}}
\end{figure*}

\begin{figure}
\includegraphics[width=0.48\textwidth]{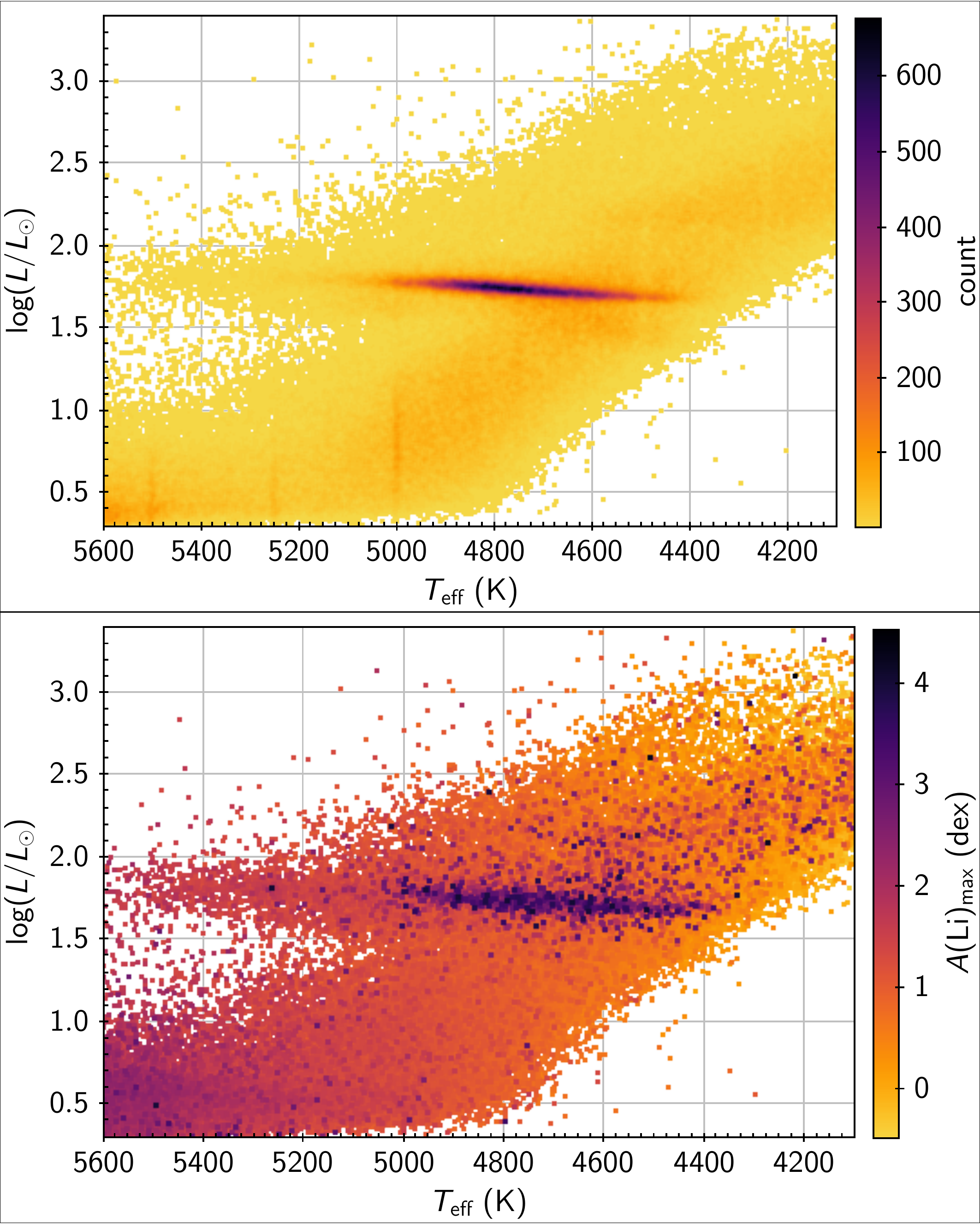}
\caption{The HR diagram for the RBS giants. Colour in the top and bottom panels represents the linear density and maximum {\it A}(Li) abundances, respectively. Note that the density visualization scale in the top panel is linear, but in Figure \ref{fig:1} scale is logarithmic.
\label{fig:LiEnrichedInHRD_WithALiInColor}}
\end{figure}

\begin{figure*}
\includegraphics[width=1\textwidth]{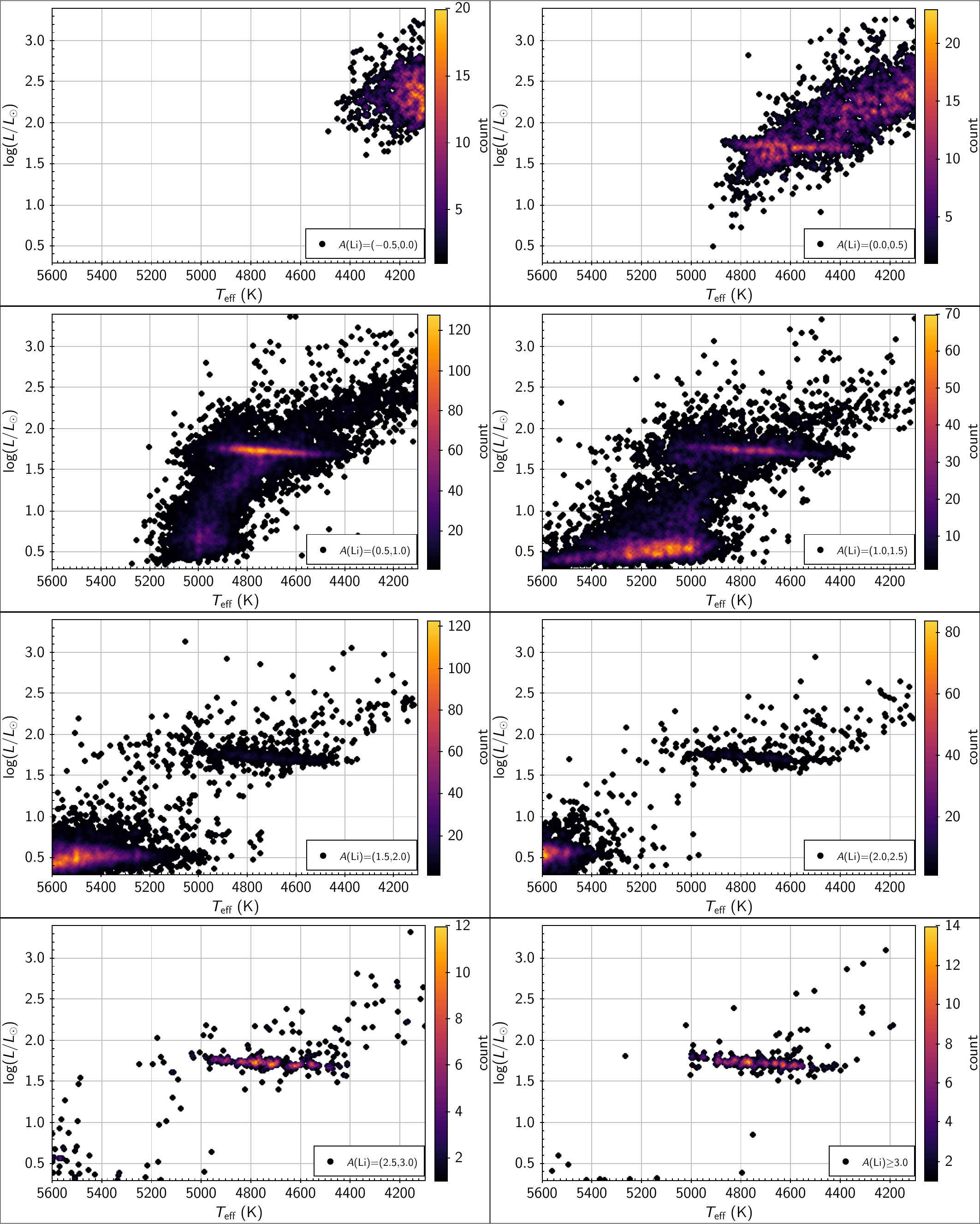}
\caption{HR diagrams for the RBS giants for Li abundances in different abundance intervals. All giants shown here have a Li detection (i.e., flag\_Li\_fe=0).
\label{fig:LiEnrichedInHRD}}
\end{figure*}

\begin{figure*}
\includegraphics[width=1\textwidth]{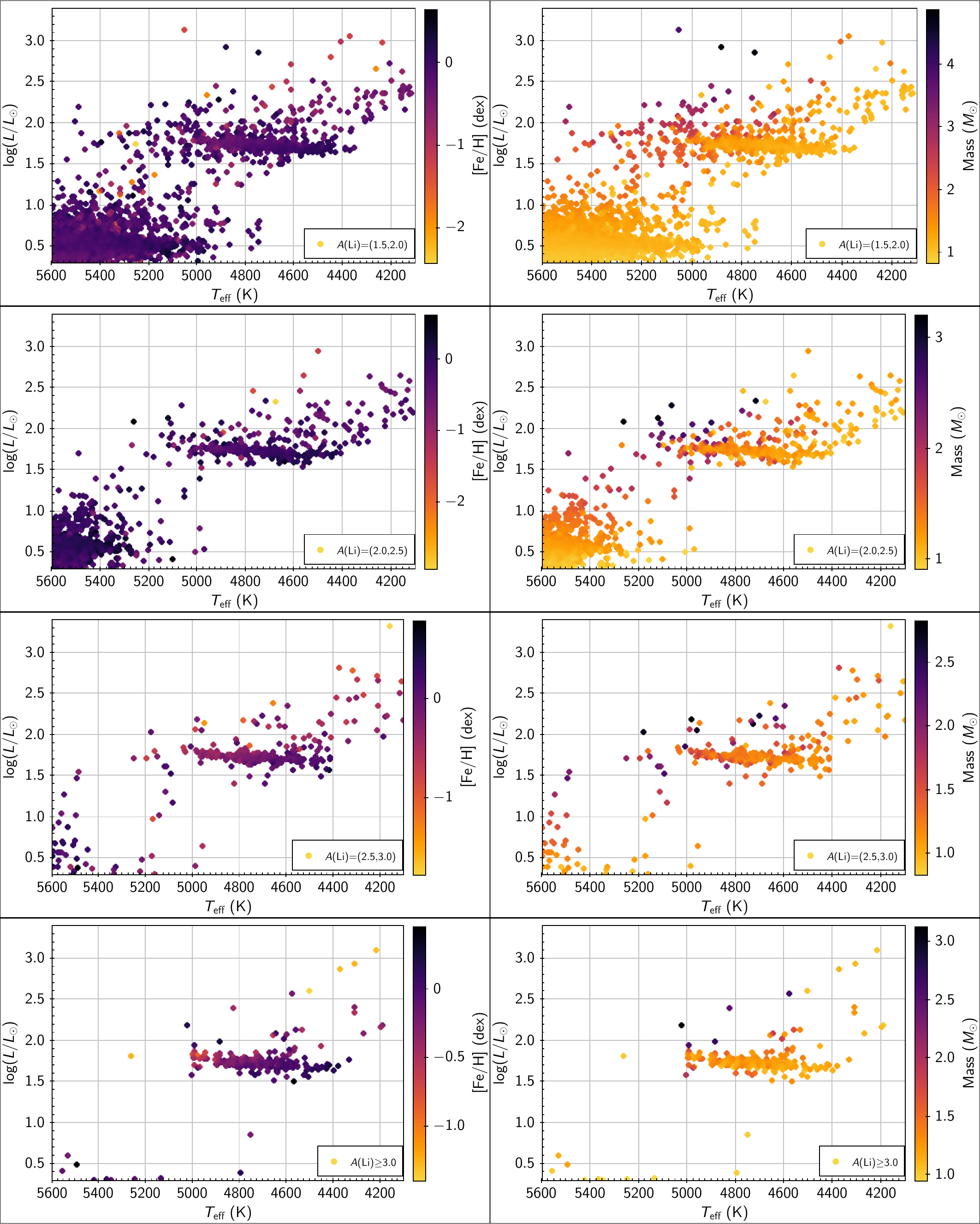}
\caption{Li abundance maps along with metallicity and mass information in colour for the RBS giants in the HR diagram. All the giants shown here have Li detections (i.e., flag\_Li\_fe=0).
\label{fig:LiEnrichedInHRD_MassFeHColor}}
\end{figure*}

\begin{figure*}
\includegraphics[width=1\textwidth]{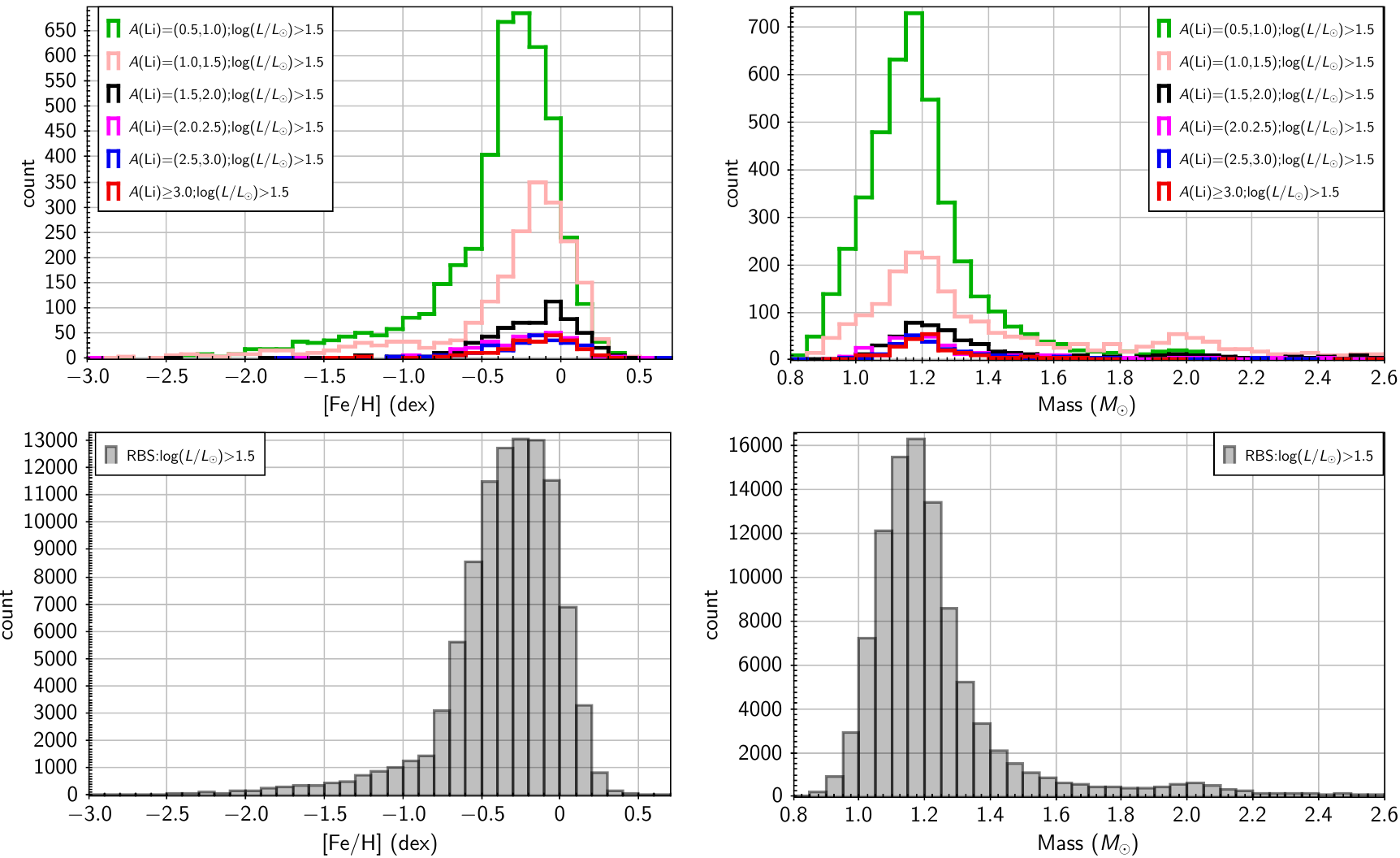}
\caption{Top row: Metallicity and mass histograms for giants plotted in Figure \ref{fig:LiEnrichedInHRD_MassFeHColor}. Histograms are shown for different Li abundance intervals, as shown in the keys in the two panels.
Bottom row: Metallicity and mass histograms for entire RBS sample with $\log (L/{\rm L_\odot}) > $ 1.5.
\label{fig:MassFeHhist}}
\end{figure*}

Our principal endeavour is to use the GALAH DR3 RBS sample (Figure \ref{fig:1}) to  (i)  explore closely the Li enrichment attributable to the He-core flash in low mass red clump giants and (ii) probe more thoroughly than before the Li enrichment predicted from thermohaline mixing at the luminosity bump on the RGB. Li-enrichment by the He-core flash was empirically proven by \cite{KumarReddy2020NatAs...4.1059K} from analysis of the GALAH DR2 Li abundances for giants but theoretical predictions of Li-enrichment are lacking apart from the recent exploration by \cite{Schwab2020ApJ...901L..18S}. In contrast to this observational--theoretical imbalance regarding Li enrichment in red clump giants, theoretical modelling of Li-enrichment at the luminosity bump \citep{EggletonDearborn2008ApJ...677..581E,LagardeCharbonnel2011A&A...536A..28L} may be more advanced than observational efforts.

To provide an intriguing glimpse into the relative roles of the two evolutionary phases of possible Li enrichment, Figure \ref{fig:HRD_and_LumALi_trends} shows in the left panel the HR diagram for the narrow range of masses $M=1.10\pm0.02\ {\rm M}_\odot$ and metallicities [Fe/H]$=0.00\pm0.10$ shown below to be rich in Li enriched giants. The middle and right-most panels show the relation between {\it A}(Li) and luminosity. The left and middle panels show giants with both Li detections and upper limits, while in the right-most panel, giants with Li detections are shown in red while giants with Li upper limits are in blue.
Gross features of the three panels are readily recognizable.
The HR diagram begins with subgiants at $\log (L/{\rm L_\odot}) \sim 0.5$ and leads into the RGB with its tip at $\log (L/{\rm L_\odot}) \simeq 3.5$. There is a steady decrease of the Li abundance {\it A}(Li) up the RGB. The LB occurs on the RGB at a luminosity just below that of the RC. At the LB, there is a noticeable increase in the counts.
Superimposed on the RGB in the HR diagram is the RC at $\log (L/{\rm L_\odot}) \simeq 1.7$ and this marks a large spread in lithium abundance up to {\it A}(Li) $\sim 4$. Apart from a few outliers at luminosities above and below the RC's luminosity, this selection does not indicate Li enrichment associated with a density increase on the RGB arising from the LB. If this selection by mass and metallicity from the RBS is representative, the outstanding message is that Li enrichment among RC giants, as \cite{KumarReddy2020NatAs...4.1059K} noted, is a striking feature. But Figure \ref{fig:HRD_and_LumALi_trends} also suggests that a search for Li enrichment associated with the LB may require subtle procedures because at the mass and metallicity ranges explored in Figure \ref{fig:HRD_and_LumALi_trends} (and at neighbouring mass and metallicities) Li enrichment from the LB appears not to be a major process.

In our effort to advance the story of Li-enrichment in giants at the red clump post-He-core flash stage and in other giants at the luminosity bump on the RGB, we combine Li abundances from the GALAH DR3 survey with stellar evolution calculations for low mass stars from the {\tt MESA} suite of computer programmes.
Our focus on possibilities of Li-enrichment in RC and LB low mass Population I giants does not mean that Li enrichment cannot occur in giants other than at the RC or LB   phases, as in mass transfer to a star from a Li-rich AGB companion (again, Li synthesis occurs via $^3$He) or in Population II giants or by other mechanisms such as engulfment of planetary bodies \citep{Alexander1967Obs....87..238A}.

\subsection{He-core burning giants - the red clump}

In Figure \ref{fig:1}, the prominent feature superimposed on the red giant branch (RGB) at $\log (L/{\rm L_\odot}) \simeq 1.7$ is the red clump (RC)  comprised of giants burning He in their core. As noted in the Introduction, \cite{KumarReddy2020NatAs...4.1059K}'s vital contribution was to show that observations demand that all RC giants in the GALAH survey were enriched in Li to differing degrees relative to stars at the tip of the RGB, the immediate predecessors of RC giants, that is Li enrichment in RC giants is `ubiquitous',  i.e., even  RC giants with the lowest Li abundance, {\it A}(Li) $\simeq -0.5$, are Li enriched. Lithium production was attributed to the He-core flash occurring at the RGB tip as the star ignites its He core. Giants with masses less than about 2.25 M$_\odot$ at solar metallicity [Fe/H] = 0 experience a He-core flash at the tip of the RGB. The upper mass limit for a He-core flash decreases slightly for metal-poor stars \citep{Karakas2010MNRAS.403.1413K}. The existence of the flash has stimulated theoretical speculation on the possibility of internal mixing and associated nucleosynthesis including production of Li from $^3$He and enrichment of the surface of the RC giant with Li. Consideration of lithium abundances among RC stars is the principal concern of our study. (Extension of the RC to hotter temperatures, as in globular clusters, marks the horizontal branch (HB), but here we use RC to refer to the entire temperature range.)

Lithium abundances among the RC giants span several dex \citep{KumarReddy2020NatAs...4.1059K}.  A simple map of  Li abundance over the HR diagram is provided in Figure \ref{fig:LiEnrichedInHRD_WithALiInColor}. The top panel shows the number density of giants across the RBS. Presence of the RC is obvious from about 4600 K to, perhaps, 5400 K  with a  peak ratio of counts in the mid-RC strip to the counts on either side of about 7. The Li map in the lower panel is colour-coded
according to the Li abundance and makes clear that the most Li-rich giants are on or maybe evolved from the RC.

More detailed maps of Li abundances are provided in the eight panels of Figure \ref{fig:LiEnrichedInHRD} showing HR diagrams for giants with {\it A}(Li) abundances in the intervals $-0.5$-0.0, 0.0-0.5, 0.5-1.0, 1.0-1.5, 1.5-2.0, 2.0-2.5, 2.5-3.0 and $>3.0$.
The full RBS  includes stars of mass from about 0.8 M$_\odot$ to 2.5 M$_\odot$.
%The full RBS sample samples stars of mass from about 0.8$M\odot$ to 2.5 M$_\odot$.
The low Li abundance panels in Figure  \ref{fig:LiEnrichedInHRD} clearly show the results of the first dredge-up (FDU) affecting subgiants and early residents on the RGB, most notably in the {\it A}(Li) = 1.0-1.5 and 1.5-2.0 panels at the luminosity $\log (L/{\rm L_\odot}) \sim 0.5$. The limit {\it A}(Li) $\geq$ 1.5 with representation across four panels corresponds approximately to the Li abundance expected following the first dredge-up, a value which has often been adopted in discussions of Li-rich giants as the lower boundary for a Li-rich giant. (Note that lithium post-FDU is further depleted in the lower mass giants -- see below.) Li abundances in these four panels include giants in which clearly Li has been enriched.
Inspection of each of these panels confirms that Li-enriched stars appear tightly confined to the RC and also to a few stars more luminous and generally cooler than the RC, i.e., to stars which may have exhausted He-burning in their core and evolved off the RC.
For each of these four samples, there is a clear gap between the RC giants and the lower luminosity RGB giants, i.e., all giants which are Li-enriched are RC giants.
There is potential confusion at luminosities above the RC between post-RC (early AGB) giants and RGB stars up to the tip of the RGB, at $\log (L/{\rm L_\odot}) \approx 3.4$.
(There may also be a possibility of confusion provided by Li-rich giants at the luminosity bump on the RGB, an issue discussed in Section \ref{sec:LiatLB}). In the panel covering the lowest Li abundances, {\it A}(Li) = $-0.5$ to $-0.0$, no stars are seen on the RC, and the majority are likely RGB giants with a luminosity below that of the RGB tip. Since further evolution to the  RGB tip may result in even lower Li abundances, RC giants emerging from the RGB tip without experiencing Li enrichment will have maximum Li abundances {\it A}(Li) $\lesssim 0.0$. But there are very few RC giants with this low Li abundance and, therefore, Li synthesis at some level from the He-core flash may occur in all giants transitioning from the RGB to the RC, as suggested by \cite{KumarReddy2020NatAs...4.1059K}.

Consideration  of the panels in  Figure  \ref{fig:LiEnrichedInHRD}  provides additional information. First, we explore the  metallicity and mass dependence of Li-rich giants \citep[see, for example,][]{DeepakLambertReddy2020MNRAS.494.1348D}. 
The metallicities (colour-coded) show that Li-rich RC giants are predominantly of near-solar metallicities (Figure \ref{fig:LiEnrichedInHRD_MassFeHColor}).
This is not a new result; see, for example, Figures 3 and 4 from \cite{DeepakLambertReddy2020MNRAS.494.1348D} and Figure 5 from \cite{MartellJeffrey2021MNRAS.505.5340M} but extends earlier results by showing that the metallicity histograms are only weakly dependent on the Li abundance.  Histograms of the metallicity for RC and (presumed) evolved RC stars in Figure \ref{fig:MassFeHhist} well illustrate that Li-rich giants are certainly representatives of the Galactic thin disk and its spread in metallicity.

Another family of histograms (Figure \ref{fig:MassFeHhist}, top-right panel) is provided for the masses of the Li-rich RC giants. As Figure \ref{fig:MassFeHhist} shows, the Li-rich giants have a narrow distribution peaking  about 1.1 M$_\odot$. The histograms extend earlier results by showing that the mass distribution is independent of Li abundance.  There is perhaps a slight increase in mass from the cool to the warm end of the RC.
%(see Figure \ref{fig:LiEnrichedInHRD_MassTeff_1}).
Distribution of Li abundances across the RBS by mass 
%in Figure \ref{fig:LumALi_RGB_MassBin}
shows Li rich RC giants from 0.8 M$_\odot$ to 2.3 M$_\odot$ with, of course, the most prominent sampling around 1.1 M$_\odot$ to 1.2 M$_\odot$. 

The histograms in the bottom panels of Figure \ref{fig:MassFeHhist} refers to the full sample covered by the RBS at luminosities $\log(L/{\rm L_\odot}) > 1.5$.
Note the large increase in counts between this sample and the histograms compiled from selected Li abundance intervals in the upper part of the figure. More importantly, note the overall similarities in the two sets of histograms. To first order, this suggests that the frequency of Li enriched RC giants is approximately independent of mass and metallicity. The two sets of histograms appear most similar for intermediate Li abundances inviting the idea that levels of Li enrichment may vary with mass and metallicity.

Close scrutiny of Figure \ref{fig:LiEnrichedInHRD} suggests several insights into Li production at the He-core flash. Most obviously, the RC's location in the HR diagram is independent of the Li abundance. For example, the mean luminosity of the  RC's core at the mean effective temperature of 4775$\pm100$ K is $\log (L/{\rm L_\odot})$ =  1.73 to within $\pm0.05$ across the bottom four panels in Figure \ref{fig:LiEnrichedInHRD}. The mild increase in luminosity with increasing effective temperature is independent of Li abundance. Similarly, the RC's width at a given effective temperature appears to be constant across these panels with the exception of a mild increase at the lowest Li abundances, which may be related to the increase in the sample size at the lowest Li abundances.  The range of effective temperatures spanned by the RC in the panels is independent of Li abundance above  {\it A}(Li)$\geq 1.5$. Also, the proportion of giants at luminosities above the RC is similar for the four panels. These characteristics about the RC and the Li abundance show that the gross properties of RC stars appear independent of the He-core flash's ability to synthesize Li.

However,  stellar evolutionary calculations fail not only to account for Li enhancement in RC giants but also do not account for their considerable spread in effective temperature at approximately constant luminosity. Such calculations predict a 1 M$_\odot$ of about solar metallicity to have approximately the observed luminosity at around 4500K. These are two failures of standard calculations, but one supposes that while their resolutions may be related, they cannot be tightly coupled because Li abundances are independent of a star's effective temperature.

It seems remarkable that the vast majority of Li enriched giants have very similar characteristics, i.e., they are or were He-core burning giants of about 1 M$_\odot$ and near-solar metallicity. Figure \ref{fig:LiEnrichedInHRD_MassFeHColor} shows that among  apparent exceptions are subgiants ($\log (L/{\rm L_\odot}) \sim 0.7$ with $T_{\rm eff} \sim 5000 -5600$ K)  whose numbers decline  with increasing {\it A}(Li). These are in all likelihood normal stars of about 1 M$_\odot$ and near-solar metallicity at different stages of an incomplete  FDU. Figure \ref{fig:LiEnrichedInHRD_MassFeHColor} also shows a few higher mass stars with  a luminosity spanning that of the RC. These too are stars observed before completion of their FDU.
Two Li enriched giants at $\sim$4800 K and at a luminosity well below that of the RC in {\it A}(Li) $\geq 3.0$ panel in Figure \ref{fig:LiEnrichedInHRD} arouse our curiosity. Both of these giants are of solar mass.
This pair ``Gaia DR2 5507464674226166144'' and ``Gaia DR2 3340856632069339776'' with {\it A}(Li) = 3.09 $\pm$ 0.11 and 3.51 $\pm$ 0.11, respectively, deserve detailed study.

In essence, setting aside few stars, our selection of  Li enriched stars from the RBS is dominated by stars of about a solar mass and a solar metallicity.

\begin{figure*}
\includegraphics[width=1\textwidth]{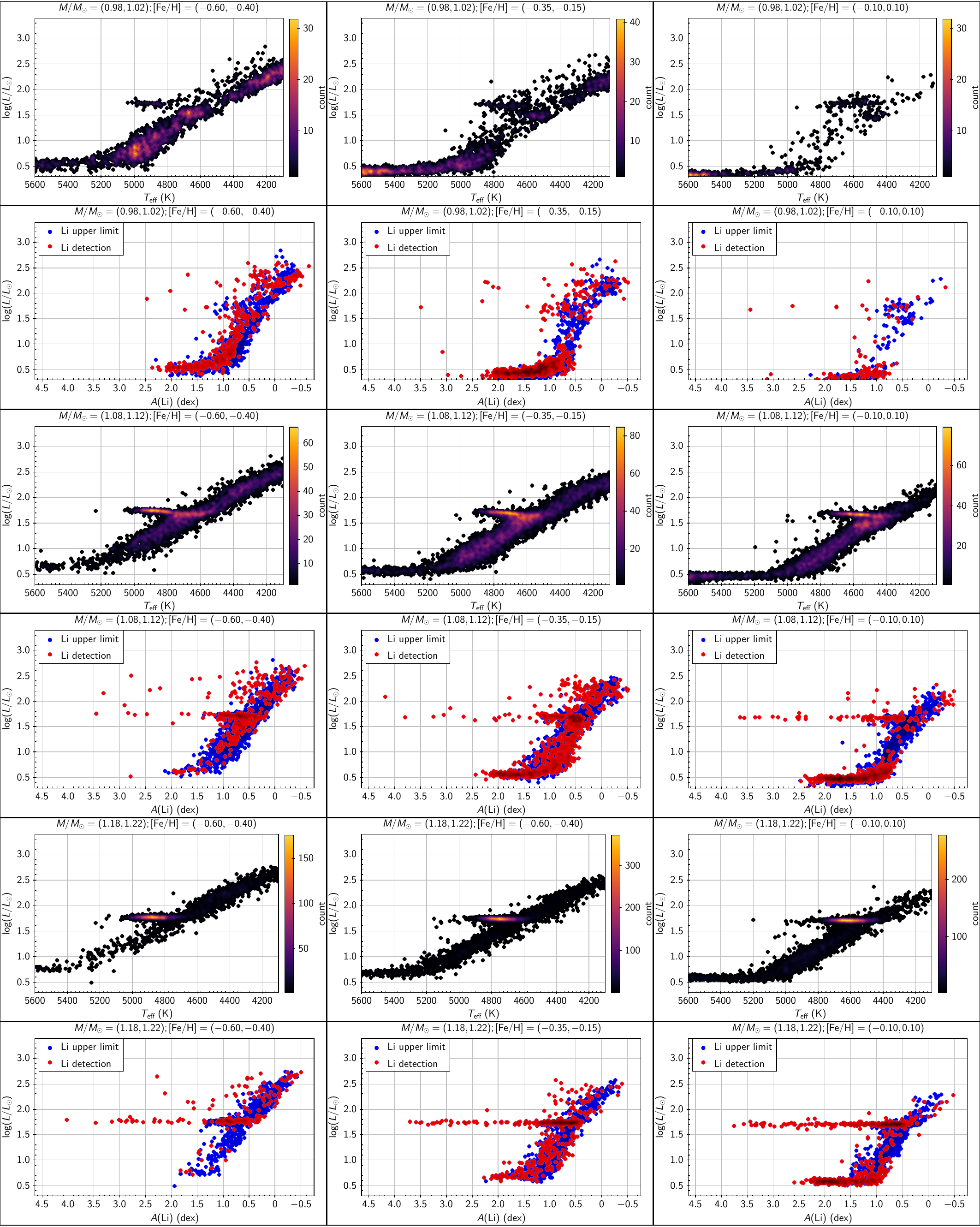}
\caption{HR diagram (odd-numbered rows) and {\it A}(Li) as a function of luminosity (even-numbered rows) for the RBS stars selected with mass and metallicity ranges as mentioned in panels' titles. In luminosity-{\it A}(Li) panels, giants with Li detections are shown in red, while giants with Li upper limits are shown in blue.
\label{fig:LB_Position_LiRich}}
\end{figure*}

\begin{figure*}
\includegraphics[width=1\textwidth]{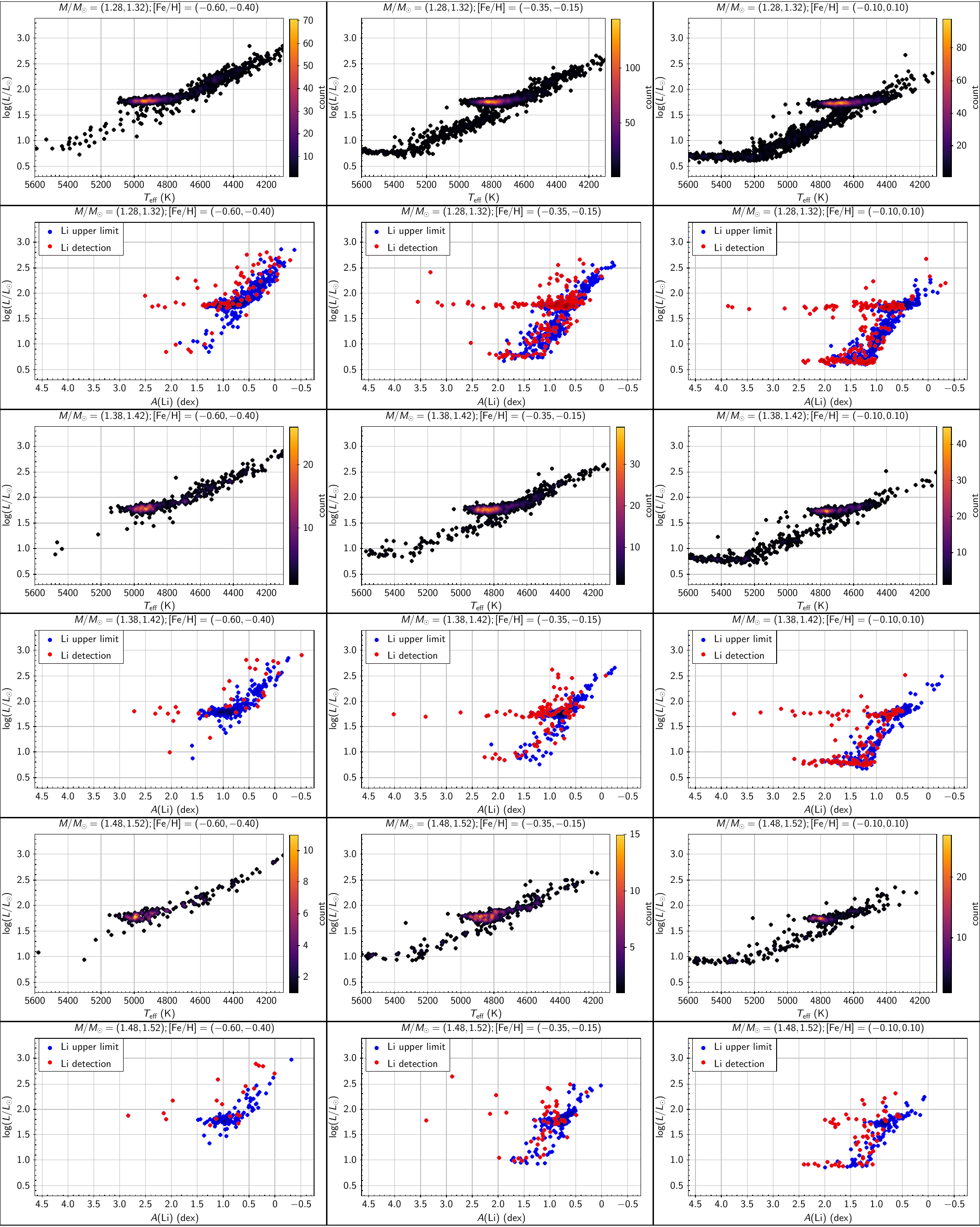}
\contcaption{
\label{fig:LB_Position_LiRich_1}}
\end{figure*}

\begin{figure*}
\includegraphics[width=1\textwidth]{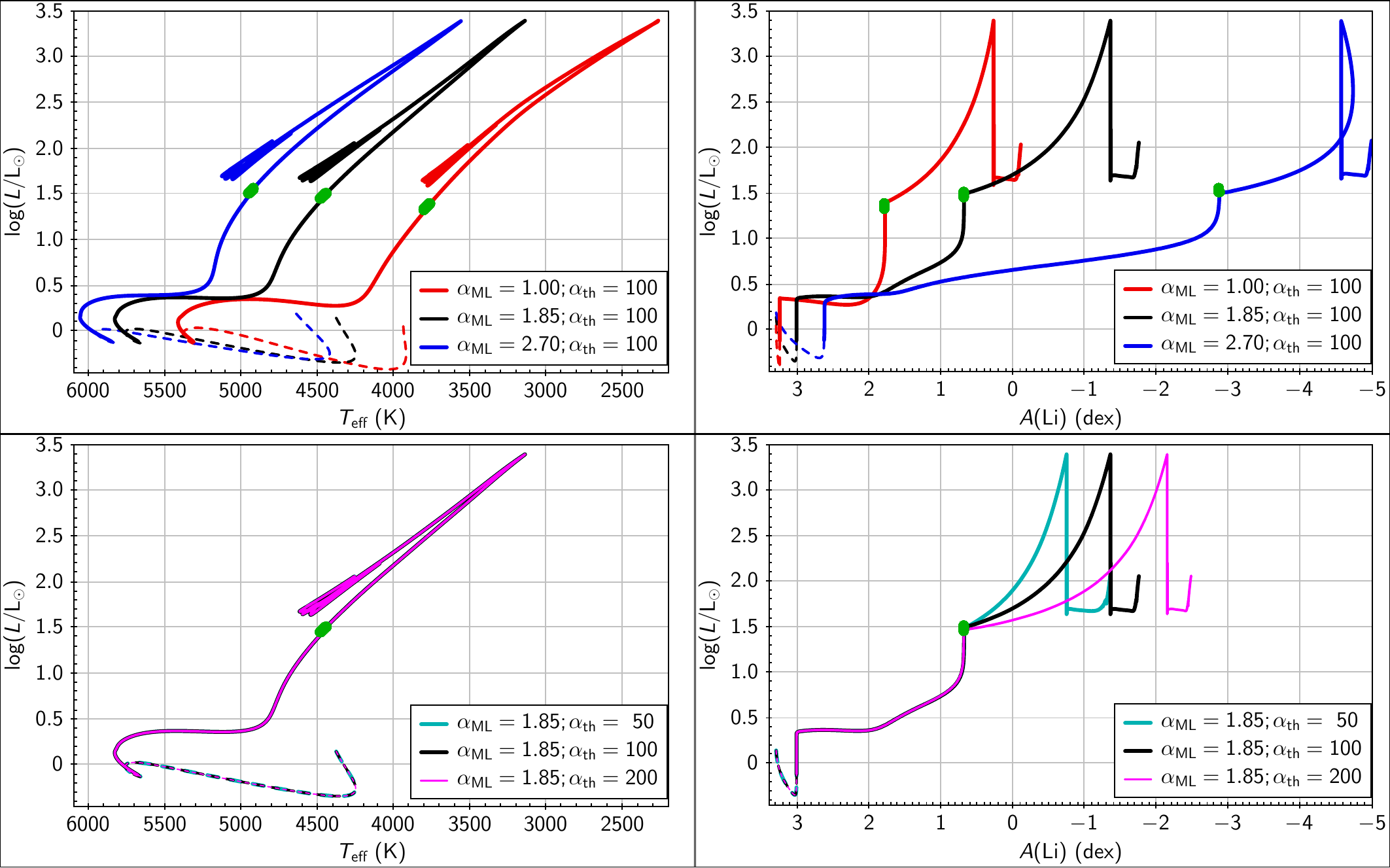}
\caption{HR diagrams and {\it A}(Li) as a function of luminosity for solar mass and solar metallicity models with different $\alpha\rm_{ML}$ and $\alpha\rm_{th}$. Top and bottom panels show a model`s sensitivity to $\alpha\rm_{ML}$ and $\alpha\rm_{th}$, respectively. Dashed lines show pre-main-sequence evolution, while solid lines show the evolution from the zero-age main sequence (ZAMS). On each track, the LB's position is marked by the green coloured blob.
\label{fig:Models_ALphaML_AlphaThm_Sensitivity}}
\end{figure*}

\begin{figure}
\includegraphics[width=0.5\textwidth]{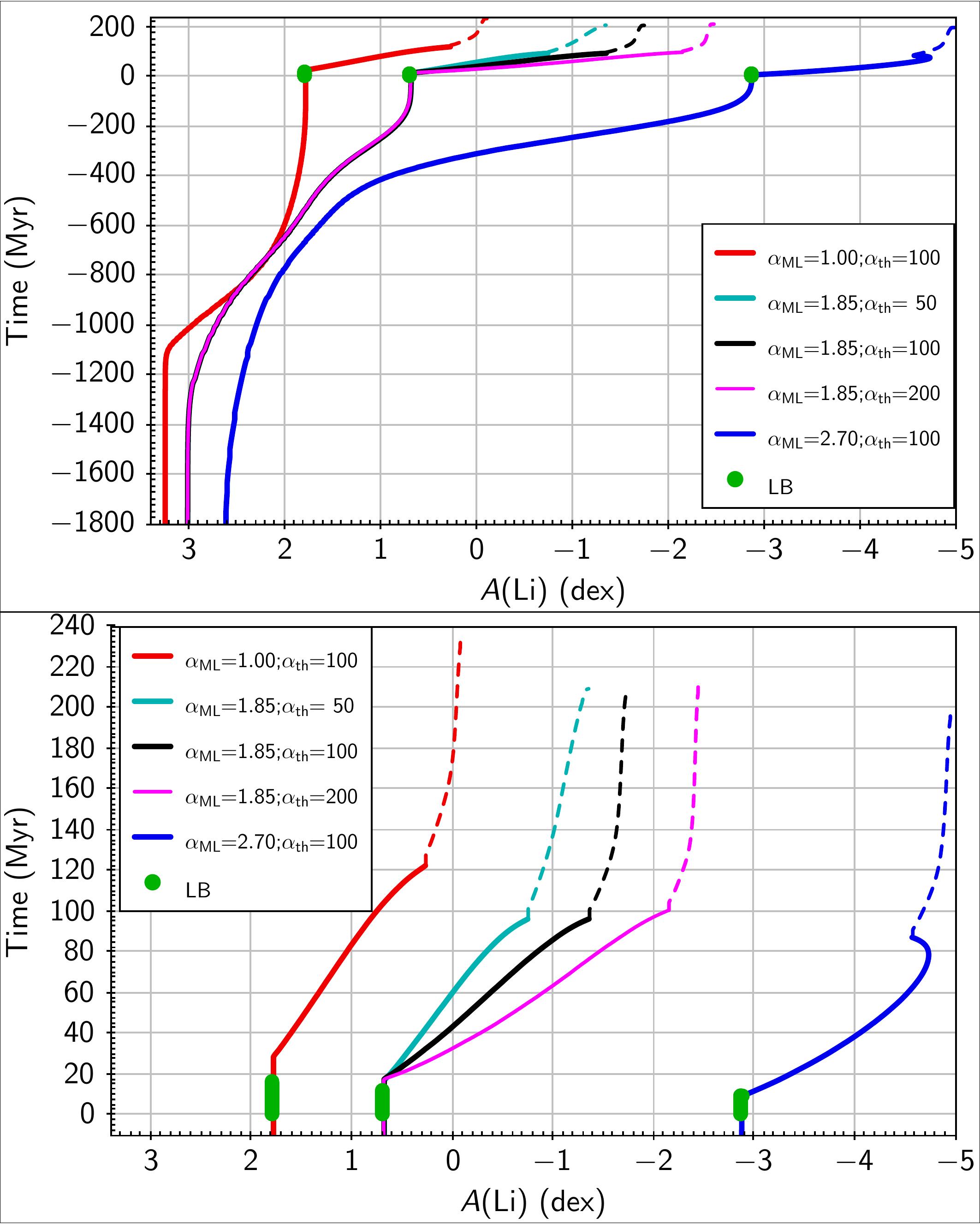}
\caption{Temporal evolution of Li for solar mass and solar metallicity models with different $\alpha\rm_{ML}$ and $\alpha\rm_{th}$. Time is shown relative to the onset of LB. Model's sensitivity to both $\alpha\rm_{ML}$ and $\alpha\rm_{th}$ is shown. The top panel covers evolution from about sub-giant phase to the CHeB phase, while the bottom panel covers evolution from near RGB on-wards. Evolution up to the RGB tip is shown as solid lines, while dashed lines show the evolution in the CHeB phase. On each track, the LB's position is marked by the green coloured blob. Timestep in each case varies from about 0.1 to 0.2 million years (Myr) during the LB, and further decreases as the star ascend the RGB.}
\label{fig:Models_ALphaML_AlphaThm_Sensitivity_Time}
\end{figure}

\begin{figure}
\includegraphics[width=0.5\textwidth]{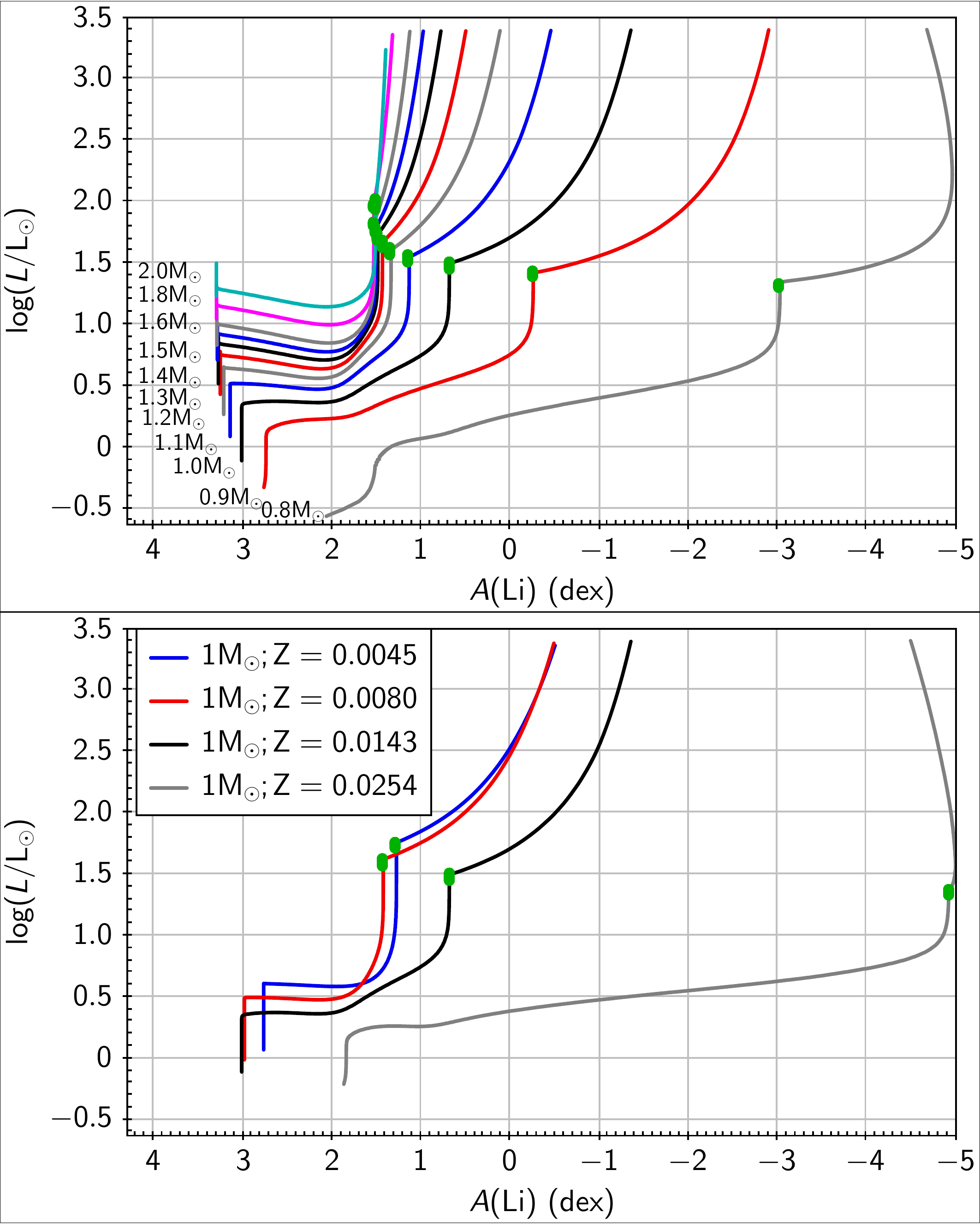}
\caption{Top panel: Predicted evolution of {\it A}(Li) as a function of luminosity for solar metallicity (Z = 0.0143) giants. Masses of 0.8 M$_\odot$ to 2.0 M$_\odot$ are shown, as labelled.
Bottom panel: Predicted evolution of {\it A}(Li) as a function of luminosity for $1{\rm M_\odot}$ mass models with metallicities Z = 0.0045, 0.0080, 0.0143, and 0.0254, which are equivalent to [Fe/H] = -0.50, -0.25, 0.00, and 0.25 dex, respectively. For all models, we use $\alpha\rm_{ML}$ = 1.85, $\alpha\rm_{th}$ = 100 and initial surface rotation $\Omega_{\it int}$/$\Omega_{\it crit}$ = 0.
For each of the case, evolution is shown from the ZAMS to the RGB tip. For all the masses, LBs are marked by green coloured blobs. The abrupt change in {\it A}(Li) occurring at about $\log (L/{\rm L_\odot}) \sim 1.5$ is attributable to the onset of thermohaline mixing at the LB.
\label{fig:ALi_Lum_Evolution_RGB}}
\end{figure}

\begin{figure}
\includegraphics[width=0.48\textwidth]{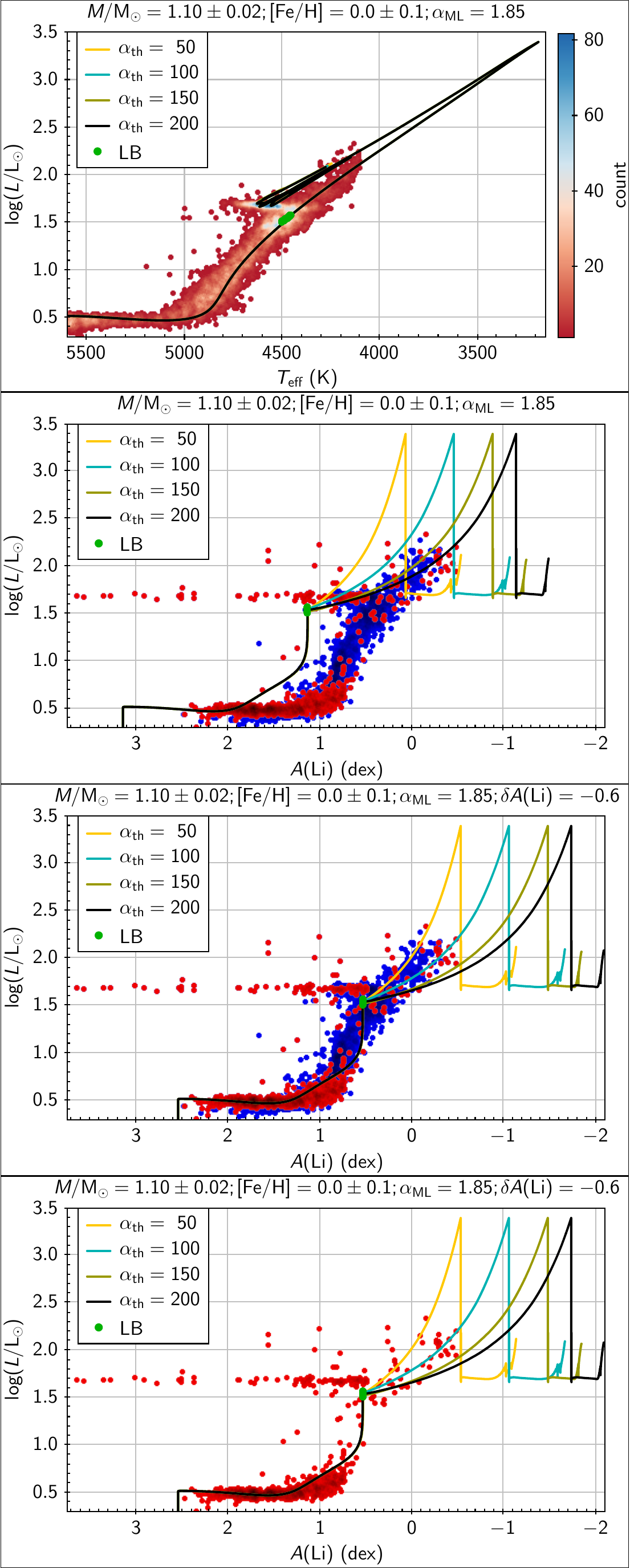}
\caption{Observed and predicted HR diagram (top panel) and {\it A}(Li) as a function of luminosity (middle and bottom panels) for RBS giants of masses 1.10 $\pm$ 0.02 M$_\odot$ and metallicity 0.0 $\pm$ 0.1 dex. Predictions are shown for best fit model with $\alpha_{\rm ML}$ = 1.85 along with $\alpha_{\rm th}$ = 50, 100, 150 and 200. The bottom panel shows luminosity-{\it A}(Li) trends after adjustment to Li abundance level. In {\it A}(Li)-luminosity panels, giants with Li detections are shown in red, while giants with Li upper limits are shown in blue.
\label{fig:HRD_ObservedAndPredicted_M1.1_Fe00}}
\end{figure}

\begin{figure}
\includegraphics[width=0.48\textwidth]{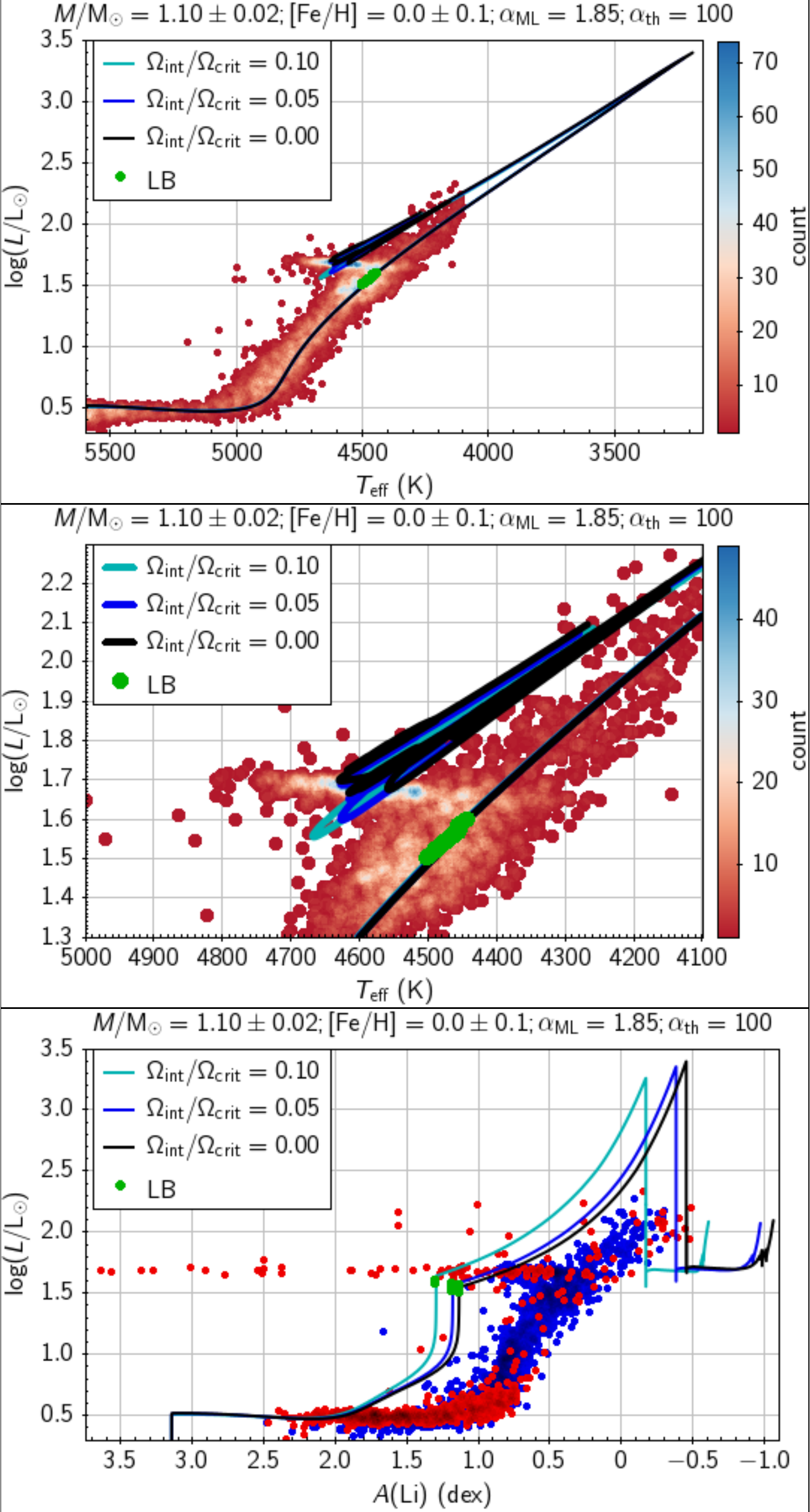}
\caption{Observed and predicted HR diagram (top panel and middle panel) and {\it A}(Li) as a function of luminosity (bottom panel) for RBS giants of masses 1.10 $\pm$ 0.02 M$_\odot$ and metallicity 0.0 $\pm$ 0.1 dex. Predictions are shown for best fit model with $\alpha_{\rm ML}$ = 1.85 and $\alpha_{\rm th}$ = 100 along with initial surface rotation $\Omega_{\rm int}/\Omega_{\rm crit}$ = 0.00, 0.05, and 0.10, which are equivalent to initial surface rotation of about 0, 15, and 30 kms$^{-1}$, respectively. In {\it A}(Li)-luminosity panels, giants with Li detections are shown in red, while giants with Li upper limits are shown in blue.
\label{fig:HRD_ObservedAndPredicted_M1.1_Fe00_Rot}}
\end{figure}

\begin{figure}
\includegraphics[width=0.49\textwidth]{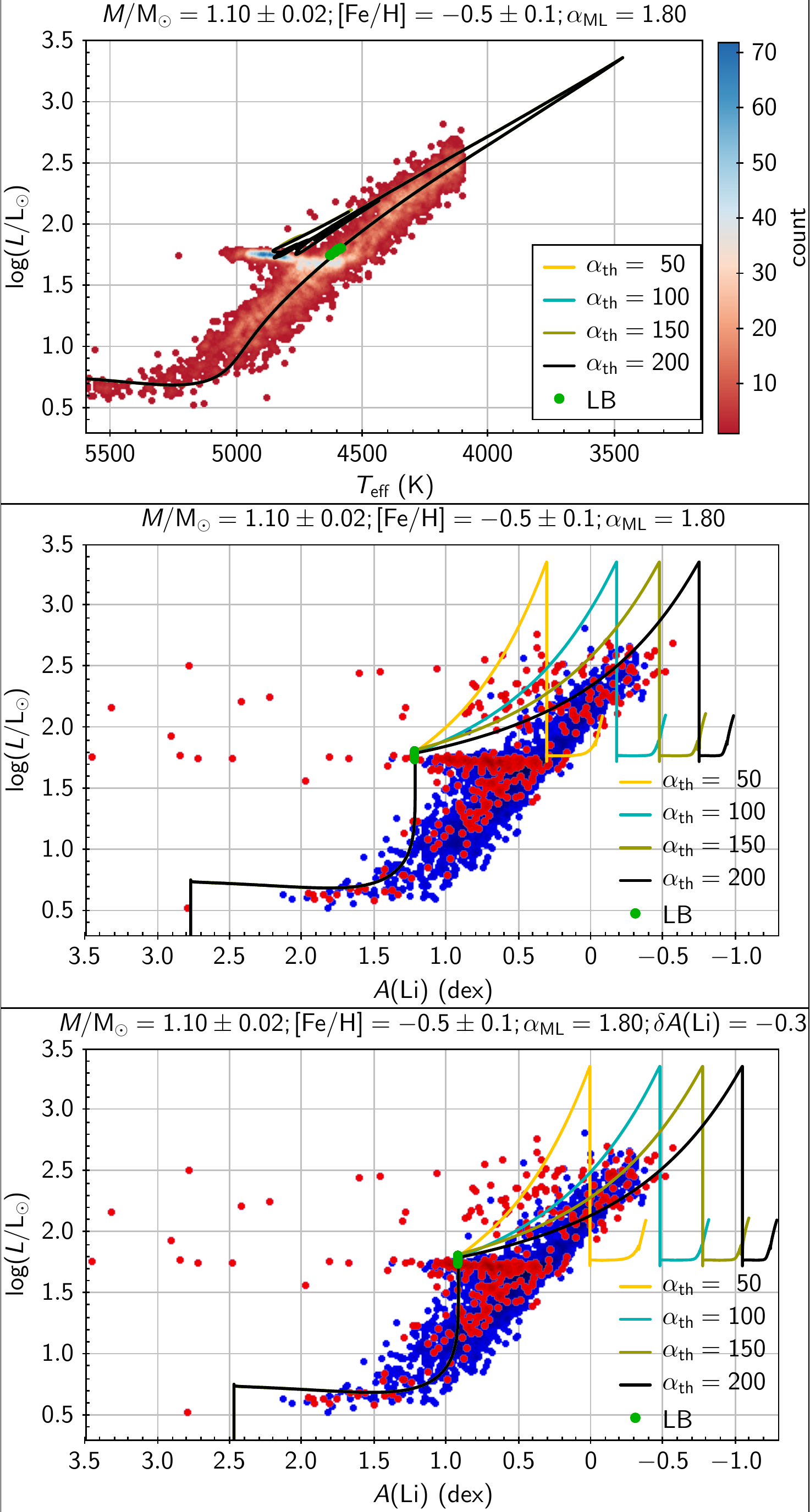}
\caption{Observed and predicted HR diagram (top panel) and {\it A}(Li) as a function of luminosity (middle and bottom panel) for RBS giants of masses 1.10 $\pm$ 0.02 M$_\odot$ and metallicity -0.50 $\pm$ 0.10 dex. Prediction are shown for best fit models with $\alpha_{\rm ML}$ = 1.80 along with $\alpha_{\rm th}$ = 50, 100, 150 and 200. In {\it A}(Li)-luminosity panels, giants with Li detections are shown in red, while giants with Li upper limits are shown in blue.
\label{fig:HRD_ObservedAndPredicted_M1.1_Fem050}}
\end{figure}

\begin{figure}
\includegraphics[width=0.49\textwidth]{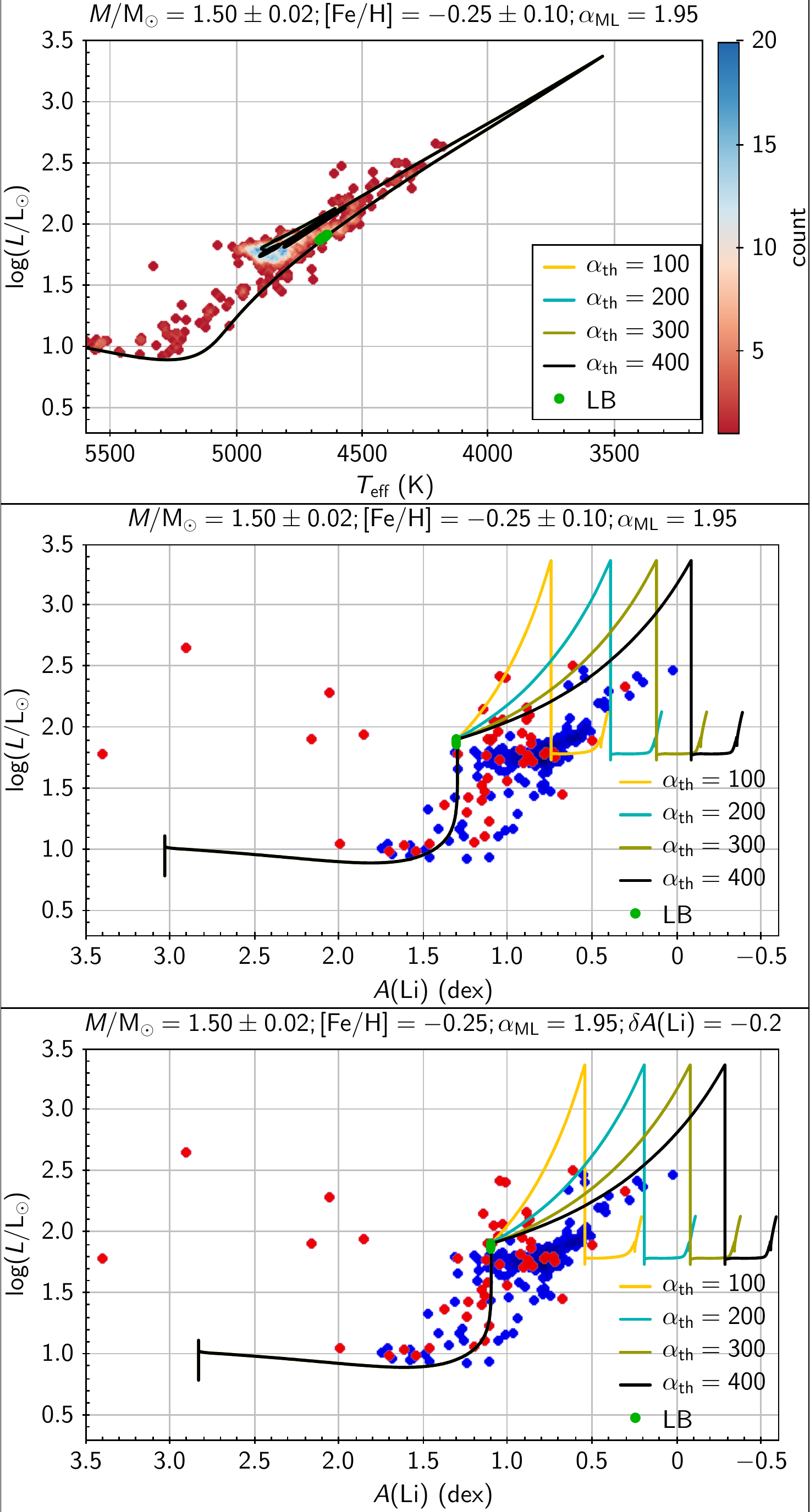}
\caption{Observed and predicted HR diagram (top panel) and {\it A}(Li) as a function of luminosity (bottom panel) for RBS giants of masses 1.50 $\pm$ 0.02 M$_\odot$ and metallicity -0.25 $\pm$ 0.10 dex. Prediction are shown for best fit model with $\alpha_{\rm ML}$ = 1.95 and $\alpha_{\rm th}$ = 100, 200, 300 and 400. The bottom panel shows luminosity-{\it A}(Li) trends after adjustment to Li abundance level. In {\it A}(Li)-luminosity panels, giants with Li detections are shown in red, while giants with Li upper limits are shown in blue.
\label{fig:HRD_ObservedAndPredicted_M1.5_Fem025}}
\end{figure}

\section{MESA modelling for low-mass stars}

To further our understanding of Li in giants, we use {\tt MESA} models to predict the Li abundance from the main sequence along the RGB  to the RC for stars in the mass and metallicity ranges covered by our RBS. Lithium dilution by the FDU and lithium evolution to the RGB's tip is modelled by {\tt MESA}.  The potential episode of Li enrichment of the atmosphere resulting from the He-core flash is not included. Also, AGB stars experiencing the third dredge-up and possible Li enrichment are neither included in selection of giants from our RBS nor in our use of the {\tt MESA} modules. 

Our initial interest in the {\tt MESA} predictions for Li abundances focussed on predicted Li abundances  for red giants up to the RGB tip and their comparison  with the observations over the mass and metallicity ranges adequately sampled by the RBS (Figure \ref{fig:LB_Position_LiRich}).
Such a comparison is also a resource with which to search for non-standard episodes of Li enrichment (or destruction) such as may occur on the RGB  or by engulfment of planetary material \citep{Alexander1967Obs....87..238A}.

\subsection{Predicted Li abundances to the RGB tip}\label{Sec:Li2RGBtip}

In pursuit of the predicted Li abundances in giants, we began with the {\tt MESA} model for a solar mass and solar metallicity model assembled by \cite{Schwab2020ApJ...901L..18S}.\footnote{\citeauthor{Schwab2020ApJ...901L..18S}'s {\tt MESA} model is publicly available on Zenodo at \url{https://doi.org/10.5281/zenodo.3960434}.}
Based on our needs, we  made some changes to \citeauthor{Schwab2020ApJ...901L..18S}'s {\tt inlist}. The prescription now includes phenomena like rotation, semi-convection, etc., along with additional resolution controls.\footnote{The source files for our solar mass and metallicity {\tt MESA} model along with the outputs for all the cases studied in this work are publicly available on Zenodo at  \url{https://doi.org/10.5281/zenodo.5086893}.}
For our calculations, we use {\tt MESA} version 15140. 
Convection in {\tt MESA} is treated using the mixing length theory (MLT) of convection, and for the current exploration we use its default formalism from \cite{CoxGiuli1968pss..book.....C}. For our models, we use the {\tt MESA}'s available recipe for thermohaline mixing adapted from \cite{Kippenhahn1980A&A....91..175K}. We also use convective premixing \citep{PaxtonSmolec2019ApJS..243...10P} which is specified by the condition {\tt use\_Ledoux\_criterion = .true.} and {\tt do\_conv\_premix = .true.} in the code. In {\tt MESA} models, the strength of the thermohaline mixing is set by a dimensionless efficiency parameter $\alpha\rm_{th}$ \citep{PaxtonCantielloMESAII_2013ApJS..208....4P}.
Thermohaline mixing is initiated when the inversion of the mean molecular weight creates the LB and the region is formally stable against convection. Mixing in the interior at this time may result in Li synthesis. This fresh Li will appear at the giant's surface if the Li enters the deep convective envelope. For cases where there is no overlap between the region of Li synthesis and the convective envelope, the surface Li abundance will be unaltered at its post-FDU value. Subsequent overlap with the convective envelope is predicted to generally reduce the surface Li abundance. 
Another parameter in {\tt MESA} modelling is the mixing length ($\alpha\rm _{ML}$) for which we generally adopt a value of about 1.85 times the pressure scale height.
Semi-convective mixing is also considered with efficiency $\alpha_{\rm sc}$ = 0.1 \citep{PaxtonCantielloMESAII_2013ApJS..208....4P}.
Our models use the standard {\tt MESA pp\_and\_cno\_extras} nuclear network, which includes 25 species and the reactions covering the pp-chains and CNO-cycles.
Our adopted {\tt MESA} nuclear reaction rates are the same as used by \cite{Schwab2020ApJ...901L..18S}.
GALAH's abundances are scaled to the solar composition from \cite{Asplund2009}. We also use the same abundance mixture and related opacity tables ({\tt a09}, {\tt a09\_co} and {\tt lowT\_fa05\_a09p} with condition {\tt use\_Type2\_opacities = .true.}) from {\tt MESA}'s {\tt kap} module.
Mass loss is not considered. A recipe for rotation is included in our models. However, in general, most of our models are non-rotating, i.e., the initial surface rotation is set at $\Omega_{\rm int}$/$\Omega_{\rm crit}$ = 0.0, and the detailed effect of rotation is discussed only for a few cases (see Section \ref{sec:ConstThermoMixing}).
Also, unlike \cite{Schwab2020ApJ...901L..18S}, our models do not include the {\it ad hoc} flash-induced diffusive mixing in the He core at the RGB tip. (However, if required, it can be easily turned on in our final prescription also.)

Selection of timestep is very crucial in the evolution of models as it helps to decide the next timestep and assist in the solution for the new structure and composition \citep[see][for more details]{PaxtonBildstenMESAI_2011ApJS..192....3P,LattanzioSiess2015MNRAS.446.2673L}. In {\tt MESA} it is primarily controlled using parameter {\tt varcontrol\_target}. In our initial test runs, we used the default minimum value of {\tt varcontrol\_target} = 10$^{-4}$ for evolution of solar metallicity models of masses 0.8 to 2.2 M$_\odot$ in 0.1 M$_\odot$ interval. Thermohaline mixing was present in all the models except for the 1.8 M$_\odot$ case. We also noticed that adoption of a slightly higher value (about three times than the default value) of {\tt varcontrol\_target} resulted in no Li reduction (i.e., missing thermohaline mixing) in the post-LB phase for masses higher than about 1.2 M$_{\odot}$ until the model has already evolved near to the tip of RGB. As a fix to this issue, we forced smaller timesteps by adopting a lower value of the parameter {\tt time\_delta\_coeff} = 0.75 than its default value of unity. Adoption of this lower timestep resulted in no significant effect on the Li trends for the cases where Li reduction near the LB was possible with a larger timestep.

Models are begun in the pre-main sequence phase at the point where the core first becomes radiative. Approach to zero-age main sequence (ZAMS) is shown by dotted lines in Figures \ref{fig:Models_ALphaML_AlphaThm_Sensitivity} and \ref{fig:ALi_Lum_Evolution_RGB}. Initial Li abundances dependent on metallicity are consistent with the Li abundances in local stars \citep{LambertReddy2004MNRAS.349..757L,DeepakReddy2020MmSAI..91..134D,RandichPasquini2020A&A...640L...1R}: {\it A}(Li) = 3.57, 3.30, 3.04 and 2.78 for [Fe/H] = $+0.25, 0.00, -0.25$ and $-0.50$, respectively. Li depletion  dependent on $\alpha_{\rm ML}$ occurs on the approach to the ZAMS: for the preferred value $\alpha_{\rm ML} \simeq 1.85$ (see below) depletion is $\sim 0.3$ and $\sim 0.15$ dex for solar metallicity stars of masses 1.0 and 1.1 M$_\odot$, respectively (Table \ref{table:1}).
Evolution is followed along the RGB and through the He-core flash at the culmination of the RGB and to He-burning in the RC  CHeB giant. We evolve all our models until the central mass fraction of $^4$He drops below 10$^{-6}$ to conclude the CHeB phase.

We first study the sensitivity of the model predictions to $\alpha\rm _{ML}$ and $\alpha\rm_{th}$. We evolved 1 M$_\odot$  and solar metallicity models with $\alpha\rm _{ML}$ = 1.00, 1.85, and 2.70, while keeping $\alpha\rm_{th}$ = 100. For $\alpha\rm_{th}$, we evolved 1 M$_\odot$  and solar metallicity models with $\alpha\rm_{th}$ = 50, 100, and 200, while keeping $\alpha\rm _{ML}$ fixed to 1.85.
The HR diagram and {\it A}(Li) trends as a function of luminosity for all these cases are shown in Figure \ref{fig:Models_ALphaML_AlphaThm_Sensitivity} from the ZAMS through the RGB tip to the end of core-He burning. The top panels show the effect of $\alpha\rm _{ML}$, and the bottom panels illustrate sensitivity to  $\alpha\rm_{th}$. An increase in $\alpha\rm_{ML}$ leads to the expected evolutionary track shifts to higher temperatures.
Higher $\alpha\rm _{ML}$ also leads to increased depletion of Li during the FDU. The temporal evolution of Li in our model stars is shown in Figure \ref{fig:Models_ALphaML_AlphaThm_Sensitivity_Time} which suggests that adoption of higher $\alpha\rm _{ML}$ leads to increased as well as prolonged depletion of Li as the star ascends the RGB. A closer look at Figure \ref{fig:Models_ALphaML_AlphaThm_Sensitivity} and \ref{fig:Models_ALphaML_AlphaThm_Sensitivity_Time} also suggests that for a given set of input parameters, models with lower $\alpha_{\rm ML}$ experience slightly delayed Li reduction at the LB in comparison to models with higher $\alpha_{\rm ML}$. Solar mass and metallicity models with $\alpha_{\rm ML}$ = 1.00, 1.85, and 2.70 starts experiencing post-LB Li reduction after about 29, 17, and 9 Myrs, respectively, from the onset of LB. However, for all three cases, the actual timestep during LB's evolution ranges from about 0.1 to 0.2 million years (Myr) and further decreases as the star evolve towards the tip of RGB. These timesteps are very small compared to the duration of LB and delays in Li reduction.
The delay in the post-LB Li reduction is sensitive to the timestep and decreases with a decrease in the timestep. A test run with smaller timestep by adopting {\tt time\_delta\_coeff} = 0.50 and 0.25 (which resulted in a timestep of about 0.1 and 0.05 Myr, respectively) for a solar mass model with $\alpha_{\rm ML}$ = 1.00 resulted in delays of about 8 and 6 Myr, respectively, along with almost tripling the computation time. It may be of interest to note that the duration of the LB depends on the assumed value of $\alpha_{\rm ML}$ and is longer for smaller values of $\alpha_{\rm ML}$. Similar is the case for overall age of the stars i.e. stars with lower $\alpha_{\rm ML}$ evolves more slowly (Figure \ref{fig:Models_ALphaML_AlphaThm_Sensitivity_Time}).
The evolutionary tracks are insensitive to $\alpha_{\rm th}$, but Li depletion commencing at the LB is, as anticipated, dependent on $\alpha_{\rm th}$.

To obtain model predictions for various masses, we evolve solar metallicity models stars of masses 0.8 to 1.6 M$_\odot$ in steps of 0.1 M$_\odot$ and 1.6 to 2.0 in steps of 0.2 M$_\odot$ from the ZAMS to the RGB tip.
The role of metallicity in the evolution of {\it A}(Li) during the RGB was explored by 1 M$_\odot$ models with  metallicities Z = 0.0045, 0.0080, 0.0143, and 0.0254, which are equivalent to [Fe/H] = -0.50, -0.25, 0.00 and 0.25 dex, respectively.
For all models, we used $\alpha\rm_{ML}$ = 1.85, $\alpha\rm_{th}$ = 100 and initial surface rotation $\Omega_{\rm int}$/$\Omega_{\rm crit}$ = 0.
Changes in the Li abundance are shown in the top panel of Figure \ref{fig:ALi_Lum_Evolution_RGB} for the series of solar metallicity models. Metallicity dependence of the Li abundance evolution is shown in the bottom panel of Figure \ref{fig:ALi_Lum_Evolution_RGB} for 1 M$_\odot$  models. The initial temperature, luminosity and $\it A$(Li) along with their values at ZAMS, at the onset of LB and at RGB tip for models in Figure \ref{fig:ALi_Lum_Evolution_RGB} are tabulated in Table \ref{table:1}.

Conditions promoting thermohaline mixing begin at a luminosity around the LB. Surface Li abundance is affected by Li destruction and formation in the interior and by the connection between the interior and the outer convective envelope.
Prescriptions for surface Li on the RGB predict in general an appreciable loss of Li beginning around the LB's luminosity, see examples in Figures \ref{fig:Models_ALphaML_AlphaThm_Sensitivity} and \ref{fig:ALi_Lum_Evolution_RGB}.
A decline of Li on the ascent of the RGB results when the convective envelope taps the region of Li synthesis. The predicted Li abundance is necessarily sensitive to the detailed modelling of thermohaline mixing, as well as stellar mass and composition. \cite{LattanzioSiess2015MNRAS.446.2673L} discuss Li abundance predictions for a 1.25 M$_\odot$ solar metallicity model given by five different codes. Across these codes, reduction of the Li abundance post-LB varies considerably for comparable mixing coefficients. An increase in {\it A}(Li) at luminosities just below the RGB tip is predicted under some conditions for some codes.
(Late (small) increases in Li abundance are seen in Figure \ref{fig:Models_ALphaML_AlphaThm_Sensitivity} for the 0.8 M$_\odot$ model with  [Fe/H] = 0 and the 1.0 M$_\odot$ models with [Fe/H] = 0.)
Such late small increases of Li will not be detectable  in DR3 and similar surveys of K giants. 
In Figure \ref{fig:Models_ALphaML_AlphaThm_Sensitivity} evolution of {\it A}(Li) is continued beyond the He-core flash to the termination of He core burning. It is at the He-core flash followed by a slight Li decrease as a RC giant. (Li synthesis at the He-core flash is not included in our models.)

A goal of our study is to confront the {\tt MESA} models with the  HR diagrams and luminosity-{\it A}(Li) trends displayed in Figure \ref{fig:LB_Position_LiRich} where in all panels, we distinguish between Li abundance measurements and upper limits. Within the constraints of {\tt MESA} models, it is apparent that the value of $\alpha_{\rm ML}$ may be determined independently of $\alpha_{\rm th}$ from an observed HR diagram. Then, determination of the best fit to the luminosity-{\it A}(Li) trend for a given mass-metallicity combination may proceed by adjustment of $\alpha_{\rm th}$. But, perhaps, the most striking observation is that the luminosity-{\it A}(Li) trend is very similar across all (M$_\odot$,[Fe/H]) pairings well sampled by the RBS, but the {\tt MESA} predictions in Figure \ref{fig:Models_ALphaML_AlphaThm_Sensitivity} and \ref{fig:ALi_Lum_Evolution_RGB} show a diverse set of luminosity-{\it A}(Li) trends.  Ahead of detailed comparisons, it is apparent that the fits of {\tt MESA} Li predictions to observations in  Figure \ref{fig:LB_Position_LiRich} will range in quality.
Notably lacking among the observed trends in Figure \ref{fig:LB_Position_LiRich} are examples matching predictions where the Li abundance is effectively constant over the RGB from completion of the FDU through  the LB's luminosity to the RGB's tip.\footnote{The mass and metallicity trends shown in Figure \ref{fig:ALi_Lum_Evolution_RGB} are qualitatively  similar to those provided by \cite{CharbonnelLagarde2010A&A...522A..10C} from the {\tt STAREVOL-GENEVA} Code.}

Investigation of  mixing on the RGB is well worthy of study because observations of Li on the RGB end at a luminosity of $\log(L/{\rm L_\odot}) \sim 2.2$ well short of the RGB tip at $\log(L/{\rm L_\odot}) \sim 3.4$. Thus, inference of the Li abundance at the RGB's tip involves a substantial extrapolation from the known Li abundances. It is the Li abundance at the tip which sets the low bound to the Li abundance among RC clump giants, i.e., the RC giants which did not experience Li addition as a result of the He-core flash.

\subsubsection{Constraining thermohaline mixing from Li predictions for red giants}\label{sec:ConstThermoMixing}

To investigate thermohaline mixing in red giants from the {\tt MESA} models, we start with the sample of giants with mass $M/{\rm M_\odot}$ = 1.10 $\pm$ 0.02 and [Fe/H] = 0.00 $\pm$ 0.10. As discussed above, the evolutionary track is sensitive to $\alpha_{\rm ML}$ and insensitive to $\alpha_{\rm th}$. A model with $\alpha_{\rm ML} = 1.85\pm0.05$ provides a good match to the observations (Figure \ref{fig:HRD_ObservedAndPredicted_M1.1_Fe00}). Li abundances for $\alpha_{\rm th} = 50, 100, 150$ and 200 were predicted. Predictions and observations are compared in the bottom panels of Figure \ref{fig:HRD_ObservedAndPredicted_M1.1_Fe00}.

Predicted {\it A}(Li) throughout the observed luminosity range is higher than the observed value. This offset, apparently independent of luminosity, may be negated by subtracting 0.6 dex from observed Li abundances, as achieved in the bottom panel of Figure \ref{fig:HRD_ObservedAndPredicted_M1.1_Fe00} showing a fair fit between predicted and observed Li abundances. The short segment of the RGB above the LB with observed Li abundances (after application of the offset) is fitted with $\alpha_{\rm th} \sim 100\pm50$.
When predictions of Li abundances are extrapolated to the RGB's tip, a Li abundance {\it A}(Li) $\simeq -1.0\pm0.3$ is obtained. Setting aside the question of the $-0.6$ offset in Li abundance, the final match of prediction to observation might seem acceptable. Two points deserve attention. First, the decline of observed Li abundances on the RGB does not clearly show the abrupt change predicted to occur at the LB's luminosity. However, as starkly shown by the bottom panel in the figure, the RGB at luminosities below the LB is poorly represented by giants with Li detections and, thus, the true slope may be ill determined.

Origins of the above  Li abundance offset may be imagined. (Also, the presence of Li upper limits among similar giants with  Li abundances  shows processes actively depleting Li are common among these stars, as giants or in earlier evolutionary phases.) Li in a main-sequence star survives in a very thin skin at the surface. Should depletion occur via the skin's base   (prior to the FDU), the Li abundance post-FDU will be reduced from its predicted value. Unless the Li-containing skin is convective, the depletion of Li will not be seen in the main sequence star yet revealed in the post-FDU giant. Similarly, mass loss from the surface of the main-sequence star may not affect its Li surface abundance but will be revealed by the FDU. There are examples of main-sequence stars with reduced Li abundances. The Sun is a notable example with an abundance of about 2.3 dex less than its initial abundance. Surely, the most extraordinary examples belong to the so-called Li-dip discovered by \cite{BoesgaardTripicco1986ApJ...302L..49B} in the Hyades at a mass of $M \sim 1.4{\rm M_\odot}$. This dip shifts to lower masses with decreasing metallicity \citep{Balachandran1990ApJ...354..310B}. \cite{ChenNissen2001A&A...371..943C} map this decline: at [Fe/H] = 0.0 and  -0.50, the mass corresponding to the dip is $M \simeq $ 1.40 M$_\odot$ and 1.26 M$_\odot$, respectively, suggesting that the Li-dip is not the primary origin for the Li offset required in Figure \ref{fig:HRD_ObservedAndPredicted_M1.1_Fe00}. Severe, even moderate, reductions of Li abundance in main sequence stars result in non-detection of Li in giants such that reductions cannot be quantified. It is curious that when well-populated with Li abundance measurements, the {\it A}(Li) - luminosity relations for the RGB in Figure \ref{fig:LB_Position_LiRich}  have a smooth outer boundary suggesting uniform initial Li abundances and effectively identical treatment of Li from the main sequence to the post-FDU giant.  The other boundary  defined by Li abundance limits from non-detection of the Li\,{\sc i} 6707 \AA\ line may be ragged and arises from stars which experienced Li loss prior to the RGB.\footnote{
It may be worth remarking that  Li with its survival  limited to a main-sequence star's outer skin differs greatly in its response to the FDU  from the C, N and O elemental and isotopic abundances  traditionally identified as measures of the FDU.}

To understand the effect of rotation, we evolved a 1.1 M$_\odot$ solar metallicity model with initial surface rotation $\Omega_{\rm int}$/$\Omega_{\rm crit}$ = 0.00, 0.05, and 0.10 which are equivalent to initial surface rotation of about 0, 15, and 30 kms$^{-1}$, respectively. A comparison of predictions with the observational data is shown in Figure \ref{fig:HRD_ObservedAndPredicted_M1.1_Fe00_Rot}. Predicted HRDs for both rotating and non-rotating model are similar except for slight disagreements at the RC and the RGB's tip. (HRDs of the rotating and non-rotating models also disagree at the main-sequence, but not shown here.) Rotating models show slightly lower luminosity and slightly higher temperature for the RC star compared to non-rotating models. This behaviour of rotating models at the RC is in contrast to observations and suggests that giants post-He-Core-flash have small or no rotation. The RGB's tip for rotating models lies at a slightly lower luminosity compared to non-rotating models (see the bottom panel of Figure \ref{fig:HRD_ObservedAndPredicted_M1.1_Fe00_Rot}). The figure suggests that rotating models experience smaller Li depletion during the FDU compared to non-rotating models. However, {\tt MESA}'s implementation of rotation-induced mixing is but one realization of this complex physical process - see  \cite{CharbonnelLagarde2010A&A...522A..10C} for an alternative prescription and different Li depletions.

Other samples in Figure \ref{fig:LB_Position_LiRich} from the RBS were studied to compare predicted with observed HR diagrams and the luminosity-{\it A}(Li) trends.
Predictions comparable to observations can be produced with small acceptable changes to $\alpha_{\rm ML}$ and $\alpha_{\rm th}$.
%As anticipated above, poorer fits between predictions and observed {\it A}(Li) are to be found among the metal-poor and the higher mass samples.
As a test of a fit to a metal-poor selection, we consider the sample for ($M$,[Fe/H]) =  ($1.10 \pm 0.02{\rm M_\odot}, -0.50 \pm 0.10$) - see Figure \ref{fig:HRD_ObservedAndPredicted_M1.1_Fem050}.
The HR diagram is well fit with a slightly lower $\alpha_{\rm ML} = 1.80 \pm 0.05$.
For this sample, the predicted luminosity of the LB falls slightly above the predicted/observed luminosity of the RC. The LB is not obvious from the observed HR diagram because it is merged with the RC.
The middle panel of Figure \ref{fig:HRD_ObservedAndPredicted_M1.1_Fem050} shows a mildly satisfactory match between observed and predicted {\it A}(Li)-luminosity trends. An additional offset of about $-0.3$ to the predicted {\it A}(Li) is required to get a  better fit. The predicted Li-luminosity trend with a sharp break in slope at the LB's luminosity seems a poor match to the observed trend, as set by giants with Li abundance detections. In contrast to Figure \ref{fig:HRD_ObservedAndPredicted_M1.1_Fe00}, Li detections span the RGB from its base to the limiting luminosity. The short segment of the RGB above the LB with observed Li abundances (after application of the offset) is fitted with a slightly higher $\alpha_{th}$ $\sim$ 150 $\pm$ 50. Extrapolation of predicted Li trends to the RGB tip gives a Li abundance {\it A}(Li) $\simeq$ -0.8 $\pm$ 0.3 at the RGB tip.

As a representative sample of higher mass giants, we consider the ($M$,[Fe/H]) = ($1.50\pm0.02{\rm M_\odot}, -0.25\pm0.10$) selection - see Figure \ref{fig:HRD_ObservedAndPredicted_M1.5_Fem025}. The rather sparse HRD  is again well fit with $\alpha_{\rm ML}$ = 1.95 $\pm$ 0.05. The LB is not prominent in the observed HR diagram. The predicted luminosity of the LB falls slightly above the predicted/observed luminosity of the RC.
Due to the small sample size, especially very few giants with a Li detection, the observed {\it A}(Li) trend is not very well defined by giants more luminous than the LB. An offset of $-0.2$ to the Li abundance level is required to match the luminosity-{\it}(Li) trend for giants less luminous than the RC. The predicted trend Li abundance with luminosity again changes slope at the LB's luminosity and may not match the smoother observed trend. The segment of giants more luminous than the RC suggests $\alpha_{\rm th} \simeq$ 300 $\pm$ 100, a value almost three times higher than required for 1.1 M$_\odot$ stars. Extrapolation of predicted Li trends to the RGB's tip gives a Li abundance {\it A}(Li) $\simeq$ $-0.1 \pm 0.3$.

Final results from detailed modeling for the above three ({\it M}, [Fe/H]) cases are summarized in Table \ref{table:2}.

\begin{figure*}
\includegraphics[width=1\textwidth]{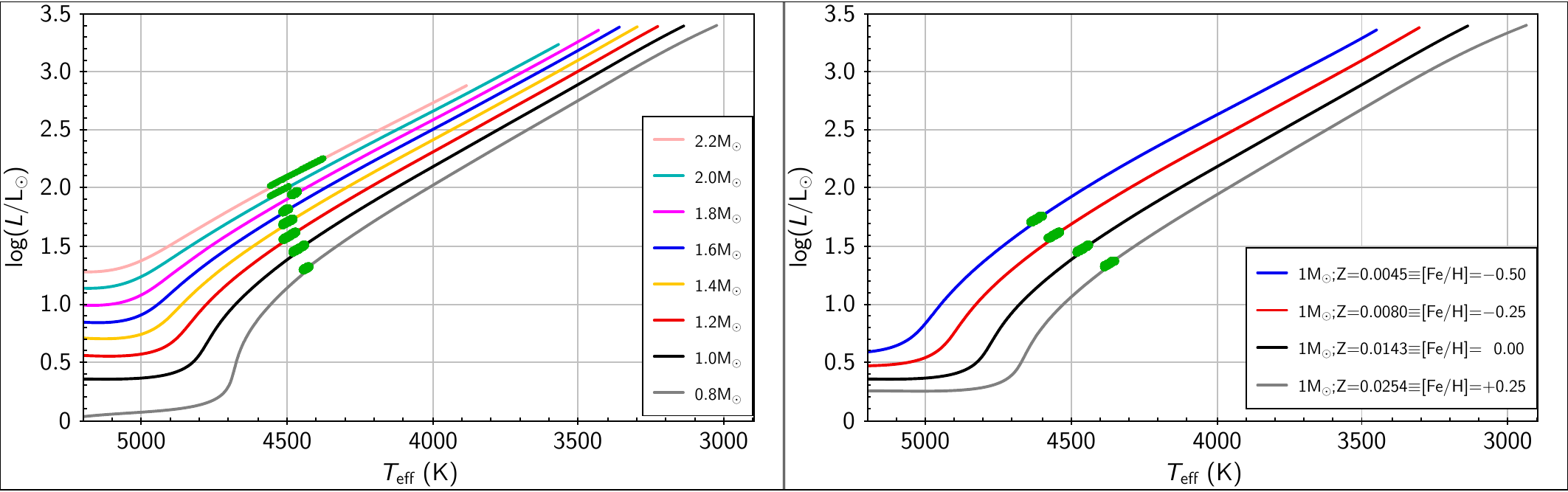}
\caption{Left panel: HR diagram showing tracks for RGB phase for solar metallicity (Z = 0.0143) models of masses from 0.8 to 2.2 in steps of 0.2 M$_\odot$. Right panel: RGB tracks for solar mass models of metallicity Z = 0.0045, 0.0080, 0.0143, and 0.0254, which is equivalent to [Fe/H] = -0.50, -0.25, 0.00 and 0.25, respectively. Shown tracks are for models with $\alpha_{\rm ML}$ = 1.85 and $\alpha_{\rm th}$ = 100. The luminous end for each of the track marks the position of RGB tip. Also, in each of the case, green blob mark the position of LB.
\label{fig:LB_Position}}
\end{figure*}

\subsection{MESA predictions for the red clump}\label{sec:MESA_RC}

\begin{figure}
\includegraphics[width=0.48\textwidth]{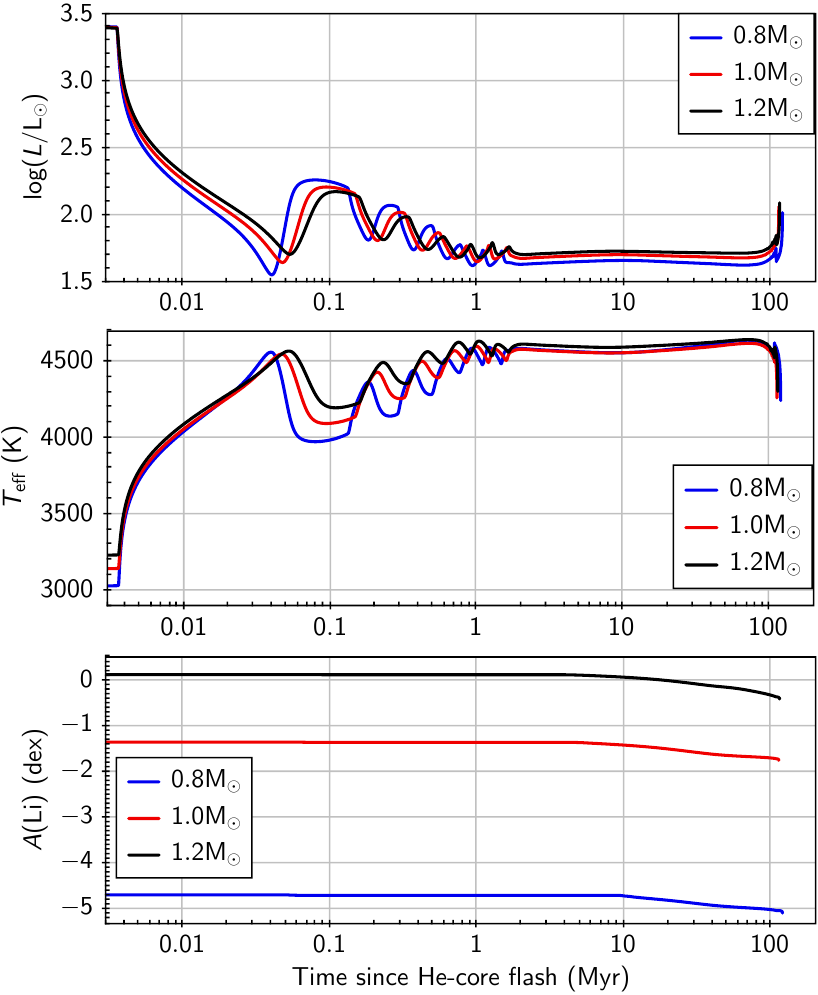}
\caption{Variation of luminosity, T$\rm_{eff}$ and {\it A}(Li) as a function of time (in million years) during the CHeB phase for solar-metallicity model stars of masses 0.8, 1.0 and 1.2 M$_\odot$ with $\alpha_{\rm ML}$ = 1.85 and $\alpha_{\rm th}$ = 100.
For the time axis, a base-10 log scale is used to highlight the fact that CHeB giants spend almost their entire life in a relatively stable state except for the brief He-flashing region, which lasts for about 1-2 Myr from the He-core flash. In the CHeB phase, giants also experience additional Li depletion. This Li depletion starts approximately about 5 Myr after the He-core flash for a solar mass and solar metallicity star.
\label{fig:CHeB_Time_EvolutionMESA}}
\end{figure}

Our MESA models include the He-core flash but not the possibility of accompanying  Li synthesis.  
In low-mass stars, there is not a single flash, but a series of He-core flashes of declining severity as He-core burning is established \citep{Thomas1967ZA.....67..420T,Despain1981ApJ...251..639D,BildstenPaxton2012ApJ...744L...6B,DeepakLambert2021MNRAS.505..642.} and the star's luminosity drops from its value at the RGB tip to the value of the RC.\footnote{\cite{IbenVol22013sepa.book.....I} provides a detailed discussion of a solar metallicity 1 M$_\odot$ model through its  He-core flashes to the third dredge-up on the AGB.}  This decline occurs so quickly that very few stars are expected to be observed in this approach to the RC. During the  RC phase, the luminosity and effective temperature are little changed for about 100 Myr for a giant of about 1 M$_\odot$. As He-core burning transitions to He-shell burning with continuing H-shell burning, luminosity increases and effective temperature decreases in the transitions from the RC to the AGB. RC evolution according to {\tt MESA} models for 0.8, 1.0 and 1.2 M$_\odot$ stars of solar metallicity is shown in Figure \ref{fig:CHeB_Time_EvolutionMESA}. Initial lithium abundances for these RC models are assumed to be the low values achieved at the tip of the RGB.
A small additional decrease in Li abundance is predicted on the RC  and in the initial ascent of the AGB (Figure \ref{fig:Models_ALphaML_AlphaThm_Sensitivity_Time} and \ref{fig:CHeB_Time_EvolutionMESA}).
An increase of the Li abundance resulting from the third dredge-up is not expected until the AGB model reaches a luminosity exceeding that of the RGB tip and an effective temperature of 3000 K or cooler. 

Notably, observations and {\tt MESA} predictions disagree sharply in two regards. First, the luminosity and effective temperature of the post He-core flash RC models are in good agreement with observations for the coolest examples in Figure \ref{fig:LiEnrichedInHRD_MassFeHColor} but cannot account for the RC's extension to higher temperatures (Figure \ref{fig:LiEnrichedInHRD_WithALiInColor} and \ref{fig:LiEnrichedInHRD}) at nearly a constant luminosity and nearly the same mass. For example, the 1 M$_\odot$ model in Figure \ref{fig:CHeB_Time_EvolutionMESA} at 10 Myr has the luminosity $\log (L/{\rm L_\odot})$ = 1.70 and $T\rm_{eff}$ = 4518 K which compares favourably with the values  for the coolest RC stars in Figure \ref{fig:LiEnrichedInHRD_MassFeHColor}.  Second, these standard {\tt MESA} models necessarily have the very low {\it A}(Li) of their RGB progenitors and cannot account for the highly variable  Li enrichment  of all - `ubiquitous' enrichment \citep{KumarReddy2020NatAs...4.1059K} --  RC stars across their temperature range. One may  speculate that these two disagreements between observations and models may have a single explanation or at least related explanations.

Lithium synthesis at the He core flash was recently investigated by \cite{Schwab2020ApJ...901L..18S} using a 1 M$_\odot$ solar metallicity {\tt MESA} model.  He demonstrated that the high Li abundance in RC stars may be explained by an event that occurs at the time of the first and strongest He-core flash. Three versions of Li synthesis and mixing events shown in \citeauthor{Schwab2020ApJ...901L..18S}'s Figure 4 give final Li abundances from about 3.7 to 1.0. The mixing introduced by Schwab is diffusive and defined by a mixing coefficient. Another parameter is the duration of the mixing event. Li may also be destroyed in a subsequent weaker He-core flash. \citeauthor{Schwab2020ApJ...901L..18S} remarks that excitation of internal gravity waves by turbulent convection during the core flash may provide the basis for the mixing, which he describes by an {\it ad hoc} mixing coefficient. These exploratory calculations do suggest that the He-core flash and attendant Li synthesis may account for varying degrees of Li enrichment seen in RC giants (Figure \ref{fig:LiEnrichedInHRD}).

\section{Lithium enrichment at the luminosity bump?}\label{sec:LiatLB}

Lithium synthesis at the He-core flash results in an abundance increase between a star at the RGB tip and the RC. In a parallel endeavour, one may search for Li abundance increases in stars at or just following the LB. Theory and observation, as discussed above, anticipate that Li abundance changes resulting from the LB are expected to be revealed subsequently as the star ascends the RGB.  In general, the Li abundance post-LB is predicted to be unchanged or to decline steadily up the RGB.  Nonetheless, the history of Li among giants suggests that one may usefully see if exceptional Li-rich giants are to be found at about the luminosity of the LB. {\tt MESA} predictions for the LB's location in the HR diagram are shown in Figure \ref{fig:LB_Position} for solar composition models with $\alpha_{\rm ML}$ = 1.8. This figure shows that the LB's luminosity increases with stellar mass and crosses the RC's luminosity at $M \simeq 1.6{\rm M_\odot}$. The LB occurs at somewhat higher luminosities and higher temperatures for metal-poor stars (Figure \ref{fig:LB_Position}). The similarity in the luminosity of RC and LB for low mass giants potentially complicates the identity of Li enriched stars in a luminosity-{\it A}(Li) diagram.

Attribution of Li enrichment directly to the LB seems possible for those (mass, metallicity) pairings providing a LB luminosity operationally distinguishable by more than 0.2 dex from the RC's luminosity.
At [Fe/H] = 0, this condition   corresponds to $M\leq 1.2 {\rm M}_\odot$ for a LB luminosity less than that of the RC or $M\leq 1.0 {\rm M}_\odot$ for a LB luminosity greater than that of the RC. At [Fe/H] $=-0.25$ these limits are reduced by about $0.2{\rm M_\odot}$.
Figure \ref{fig:LB_Position_LiRich} shows that only the search among the samples with a luminosity less than that of the RC is practical for the DR3 survey; far too few giants with $M\geq 1.6 {\rm M}_\odot$ are included in DR3.
 
One suitable sample is shown in Figure \ref{fig:HRD_and_LumALi_trends}. The distinct locations of the RC and LB are well shown. Surely, the outstanding feature in the luminosity- {\it A}(Li) diagram is the wide range of Li abundances at the RC's luminosity. At the LB's luminosity, there are no giants with an obvious Li enrichment; most of the Li abundances are upper limits. Several panels in Figure \ref{fig:LB_Position_LiRich} confirm this result for other (mass,metallicity) combinations.

Figure \ref{fig:LB_Position_LiRich} shows that for all (mass, metallicity) combinations, the Li abundance post-LB declines at a similar rate as the giant evolves beyond the LB. If significant Li enrichment were to occur at the LB, it may vary even between stars with identical properties, as it does among RC giants. Then, the width of the {\it A}(Li) -- $\log (L/{\rm L_\odot})$ trend on the RGB above the LB would be increased relative to the width below the LB. In principle, this test of Li enrichment at the LB seems applicable to all (mass, metallicity) combinations provided that the RGB is well populated. Unfortunately, the presence of many upper limits to the Li abundances impairs the test. In addition, at luminosities above the RC and LB, Li enriched RGB stars are difficult to isolate from evolved Li enriched RC giants. 

Our conclusion that Li-rich giants do not occur on the RGB {\bf at}  the LB differs from remarks made by \cite{MartellJeffrey2021MNRAS.505.5340M} from their analysis of Li-rich giants in the GALAH DR3 survey. They conclude that lithium enrichment is not limited to RC giants but is also present among an appreciable number of RGB giants: ``red clump stars are 2.5 times as likely to be lithium-rich as red giant branch stars.''  This disagreement with our analysis may arise from \citeauthor{MartellJeffrey2021MNRAS.505.5340M}'s use of Kiel diagrams ($T_{\rm eff},\log g)$ where giants are assigned to either the RGB or the RC using the  Bayesian probability estimates provided in column {\tt is\_redclump\_bstep}. Giants with {\tt is\_redclump\_bstep} $\geq$ 0.5 and {\it WISE} $W_{\rm 2}$ absolute magnitude in the range $|W_{\rm 2}$ $+$ 1.63$|$ $\leq$ 0.80 are classified as RC giants, while giants with {\tt is\_redclump\_bstep} $<$ 0.5 or absolute magnitude outside the range $|W_{\rm 2}$ $+$ 1.63$|$ $\leq$ 0.80 are classified as RGB giants.
This results in a classification of all the luminous giants with $W_{\rm 2}$ $<$ -2.43 or $log(L/{\rm L_\odot})$ $ \gtrapprox$ 2.1 as RGB giants, which in actuality are expected to be a mixture of RGB, early CHeB, and post-RC giants.
The use of Bayesian estimates, which vary with luminosity, effective temperature and metallicity, appear to have introduced some intermixing of the RGB and RC populations. We prefer to assign evolutionary phases using a giant’s location in the HR diagram ($\log(L/{\rm L_\odot}), T_{\rm eff}$) with respect to the appropriate evolutionary track. As noted above, there are regions in the HR diagram where ambiguities may arise depending on the accuracy of the parameters used to place a giant on the HR diagram, for example, at the confluence of the RC and the RGB.

In summary, Li enrichment {\it at}  the LB is not detected in the giants well represented by DR3. An upper limit to the Li abundance increase of about 0.3 dex is suggested by visual inspection of Figure \ref{fig:LB_Position_LiRich}. This may provide a novel challenge to the theoretical modelling of thermohaline mixing at the LB.

\section{Compositions of red clump giants}

The chemical composition of Li enriched giants has been studied by \cite{DeepakLambertReddy2020MNRAS.494.1348D} to a great extent based on the GALAH DR2 and then by \cite{MartellJeffrey2021MNRAS.505.5340M} based on data from the GALAH DR3. The conclusion is clear: compositions of Li-rich giants are not distinguishable from other giants. Here, attention is returned specifically to the RC giants. All of these stars have been affected by the He-core flash, which led to alterations of the Li abundance by an amount varying from small to substantial. In examination of the [X/Fe] versus [Fe/H] plots, the emphasis here is on elements affected by He-burning and $s$-processing. The earlier conclusion about compositions of Li enriched and normal giants is confirmed.

Unfortunately, elements provided by the GALAH survey are not well suited to probing surface contamination by products of He burning mixed to the surface during the short-lived episode of He-core flashes or the longer period of He-core burning.
Just 229 giants in the RC luminosity strip (with $\log (L/\rm L_\odot)$ of 1.6-1.9) have a determined C abundance.
%Just four Li-rich RC giants have a determined C abundance.
Nitrogen, which might be affected via associated H-burning CN-cycling, is not provided in the GALAH survey. Oxygen, which is well represented in this survey, shows the [O/Fe] versus [Fe/H] relation expected for thin-disk giants with no alterations dependent on Li abundance. This suggests that, although internal He-burning may have altered surface Li abundances,  oxygen which is possibly a direct product of He-burning has not been altered.

Coupling of $s$-processing with He-burning is a theoretical possibility.  GALAH provides 11 elements from Rb to Sm as at least partial monitors of an $s$-process and Eu as a measure of the $r$-process.  For four elements - Rb, Sr, Mo and Ru - abundances are sparsely reported. Among the RC giants, there is no evidence that abundances of these  11 elements differ according to their Li abundances or differ at all from the [X/Fe] versus [Fe/H] trends expected for the thin disk.\footnote{The [X/Fe] versus [Fe/H] for X = Sr, Mo and Ru show an odd bifurcation. A majority of the measurements provide [X/Fe] $\simeq 0$, as found for other $s$-process dominated elements, but a minority of giants suggest a second trend of increasing [X/Fe] with decreasing [Fe/H]  attaining [X/Fe] $\sim +1$ at [Fe/H] $\sim -0.4$. Is this exceptional behaviour a reflection of systematic errors, perhaps linked to misidentification of lines? }

In brief, elements examined across the RBS sample of RC giants show no abundance anomalies except for Li, but this conclusion can not yet be extended to principal products of He-burning (C and N and their isotopes)  for lack of observations.

\section{Concluding remarks}\label{sec:conclusions}

Since their discovery in 1982 by \citeauthor{WallersteinSneden1982}, Li-rich giants have occasioned considerable attention by both observers and theoreticians searching for the evolutionary phases at which a stellar atmosphere's Li abundance may be increased by up to several dex. For red giants considered as single stars, the search for internal lithium production has been directed at two evolutionary phases in low mass giants (say, $M \lesssim 2 {\rm M_\odot}$): the luminosity bump on the RGB and the He-core flash.  It is now clear that an overwhelming majority of lithium-rich red giants are He-core burning giants, members of the so-called red clump, featuring lithium synthesized by the Cameron-Fowler mechanism during (presumably) the He-core flashes occurring in the RC star's predecessor at the RGB's tip. Detectable Li synthesis is not detectable at the luminosity bump. The preeminence of the idea that the majority of lithium-rich giants are He-core burners ends a long debate about the relative roles of the red clump and the luminosity bump. (The role of binary stars in creating Li-rich giants has been examined. \cite{CaseyHoNess2019ApJ...880..125C} proposed that Li was synthesized in a RGB giant when  Li synthesis was triggered by perturbations arising from a companion star. The observation that the binary frequency for Li-rich giants resembles that for Li normal giants \cite{JorissenVanWinckel2020A&A...639A...7J} is a challenge to the universal viability of this binary hypothesis.)

The conclusion about the dominance of the He-core flash in creating Li-rich giants was strengthened by Kumar et al.'s (\citeyear{KumarReddy2020NatAs...4.1059K}) use of the GALAH DR2  survey.  \citeauthor{KumarReddy2020NatAs...4.1059K} noted that the population of the Li-rich population was at the luminosity of the red clump with Li abundances {\it  A}(Li) spanning from about $-0.5$ to $+3.7$. Red clump residents have evolved from the RGB and most likely from a star at the RGB's luminosity tip. Thus, the lithium abundance at the RGB's tip provides the baseline for a clump giant. \citeauthor{KumarReddy2020NatAs...4.1059K} inferred by extrapolation of the GALAH abundances for the RGB that the abundance at the tip was {\it A}(Li) $\sim -0.9$. Thus, this baseline and the Li abundance range $-0.5$ to $+3.7$ for  RC  giants led \citeauthor{KumarReddy2020NatAs...4.1059K} to a conclusion that all RC residents were enriched in lithium: across the range of Li abundances,  the enrichment is slight to dramatic. \citeauthor{KumarReddy2020NatAs...4.1059K} referred to lithium enrichment as `ubiquitous' among red clump giants. Our probing of red clump giants in the GALAH DR3 endorses \citeauthor{KumarReddy2020NatAs...4.1059K}'s conclusions about the ubiquity of lithium enrichment among RC giants.

{\tt MESA} models do not fully account for the run of Li abundance along the RGB. Across the sample examined here, the maximum predicted Li abundances of RGB giants are about 0.2 - 0.6 dex greater than observed, which suggests a loss of Li within the main sequence star prior to the RGB. Loss of Li is also suggested by the absence of detectable Li in many RGB giants. All mass-metallicity samples of RGB giants exhibit a  smooth decrease of Li abundance with increasing luminosity, but the {\tt MESA} predictions provide for a steeper decline of Li abundance at luminosities above that of the LB.  A possible resolution of this discrepancy is to invoke missing  `physics' in the models, such as rotation-induced mixing.

Our analysis  shows that the lithium enriched population peaks at a mass $M \simeq$ 1.1 M$_\odot$ and metallicity [Fe/H] $\simeq -0.2$. These values reflect the mass and metallicity distributions of the GALAH population as a whole. The lithium richest and lithium poorest stars in a mass-metallicity combination may appear at the same effective temperature and at the same luminosity. The ubiquity of Li enrichment among RC giants demands redefinition and restatement of the frequency of Li-rich giants among a population, previously stated to be at the few per cent level. If ubiquity is the correct description for RC giants, all such giants with mass $M \simeq$ 1.1 M$_\odot$ and metallicity [Fe/H] $\simeq -0.2$ have experienced the addition of fresh Li to their atmosphere. Future investigations exploring red giants beyond the masses and metallicities well covered by the RBS are awaited with interest.

MESA models fail to account for observations of RC giants. (Li synthesis was not included in our models.)  There is an observed spread in effective temperature (and Li abundance) for giants of a given mass and metallicity. But models predict He-core burning to occur at essentially a single effective temperature before the star evolves to the AGB phase at lower temperatures and higher luminosity.  This observed spread in effective temperature and an uncorrelated spread in Li abundance suggests that the sequence of He-core flashes initiating the RC phase at the RGB's tip may result in internal adjustments not considered by {\tt MESA} (and similar models). The primary search for the origin for the effective temperature spread among red clump giants must be led by theoreticians. This issue, however,  may be explored by examining compositions of red clump giants across their full spread in effective temperature. No abundance anomalies are shown by the GALAH compositions, but key nuclides are very incompletely reported (as C) or not reported at all (as N).
 
The luminosity bump has been offered as an opportunity to enhance the surface lithium abundance. Guided by the {\tt MESA} predictions on the bump's location, we searched for a direct association with Li enriched giants. Across the mass and metallicity combinations providing HRDs with well-populated RGBs, we found no evidence for the presence of Li enriched giants at the LB: an upper limit of 0.3 dex to the increase in Li abundance is set. This provides an interesting limit with which to test calculations of lithium synthesis occurring at the luminosity bump. Such synthesis does directly affect the Li abundance development on the RGB above the LB, where without obvious exception, the Li abundances are less than at pre-LB RGB values. 
 
The saga of the origins of lithium enrichment in low mass red giants has now transitioned to the challenge of providing a clear picture of Li synthesis at the  He-core flashes in luminous RGB giants.  Clarification of the identification of Li enriched giants is a notable achievement, but questions concerning the quantitative accounting for giants' lithium abundances remain, most especially difficult questions about the He-core flashes and their effect on the structure of red clump giants which may in many cases depart markedly from the clump giants described by standard evolution calculations. Although direct observations of Li abundance at the LB have not been found, observations of Li and light nuclides (C, N and O) affected by thermohaline and other forms of mixing along the full RGB deserve pursuit before the understanding of RGB giants may be deemed complete.  Undoubtedly, as the history of lithium in observational astronomy reminds us, other puzzles about the lithium abundances among giants as single and binary stars will occur unexpectedly.

\section*{Acknowledgments} \label{sec:acknowledgments}

We are indebted to the anonymous reviewer and to John Lattanzio for helpful correspondence and, most especially, to the GALAH team and their publication of the data release DR3. Without the catalogues associated with DR3, our analysis would have been impossible. A similar debt is due to Bill Paxton and his colleagues for developing the {\tt MESA} suite of programmes. This work has also made use of NASA's Astrophysics Data System.
The data for the GALAH survey are acquired through the Australian Astronomical Observatory, under programs: A/2013B/13 (The GALAH pilot survey); A/2014A/25, A/2015A/19, A2017A/18 (The GALAH survey phase 1); A2018A/18 (Open clusters with HERMES); A2019A/1 (Hierarchical star formation in Ori OB1); A2019A/15 (The GALAH survey phase 2); A/2015B/19, A/2016A/22, A/2016B/10, A/2017B/16, A/2018B/15 (The HERMES-TESS program); and A/2015A/3, A/2015B/1, A/2015B/19, A/2016A/22, A/2016B/12, A/2017A/14 (The HERMES K2-follow-up program). We acknowledge the traditional owners of the land on which the AAT stands, the Gamilaraay people, and pay our respects to elders past and present. This paper includes data that has been provided by AAO Data Central (\url{https://datacentral.org.au/}). We also acknowledge using the Delphinus and Fornax server at IIA, Bengaluru, to evolve our model stars.

\section*{ORCID iDs}
Deepak: \url{https://orcid.org/0000-0003-2048-9870}\\
David L. Lambert: \url{https://orcid.org/0000-0003-1814-3379}

\section*{Data Availability}

The observational data underlying this article are publicly available in the Data Central at \url{https://cloud.datacentral.org.au/teamdata/GALAH/public/GALAH_DR3/}.
The source files for our solar mass and metallicity {\tt MESA} model along with the outputs for all the cases studied in this work are publicly available on Zenodo at \url{https://doi.org/10.5281/zenodo.5086893}.

\bibliographystyle{mnras}
\bibliography{ref} % if your bibtex file is called ref.bib

%%%%%%%%%%%%%%%%% APPENDICES %%%%%%%%%%%%%%%%%%%%%
\appendix
%$$$$$$$$$$$$$$$$$$$$$$$$$$$$$$$$

%\section{Extra material}
\section{Tables summarizing results from the {\tt MESA} models}

\begin{table*}
\caption{The initial temperature, luminosity and $\it A$(Li) along with their values at ZAMS, at the onset of LB and at the tip of the RGB for the models shown in Figure \ref{fig:ALi_Lum_Evolution_RGB} and \ref{fig:LB_Position}.\label{table:1}}
\footnotesize
\begin{tabular}{cc|cccc|cccc|cccc}
    \hline
$M$/M$_\odot$ & Z & \multicolumn{4}{c|}{Temperature (K) at} & \multicolumn{4}{c|}{$\log$ ({\it L}/L$_\odot$) at} & \multicolumn{4}{c}{{\it A}(Li) at}\\
 & & PMS & ZAMS & LB & RGB-tip & PMS & ZAMS & LB & RGB-tip & PMS & ZAMS & LB & RGB-tip\\
\hline
0.8 & 0.0143 & 4082 & 4894 & 4422 & 3028 & -0.19 & -0.56 & 1.32 & 3.40 & 3.30 & 2.07 & -3.03 & -4.67\\
0.9 & 0.0143 & 4243 & 5323 & 4430 & 3087 & -0.01 & -0.33 & 1.42 & 3.40 & 3.30 & 2.77 & -0.26 & -2.90\\
1.0 & 0.0045 & 4681 & 6165 & 4600 & 3454 & 0.36 & 0.07 & 1.76 & 3.36 & 2.78 & 2.77 & 1.28 & -0.50\\
1.0 & 0.0080 & 4533 & 5936 & 4537 & 3307 & 0.26 & -0.01 & 1.62 & 3.38 & 3.04 & 2.99 & 1.43 & -0.49\\
1.0 & 0.0143 & 4380 & 5678 & 4438 & 3141 & 0.15 & -0.11 & 1.51 & 3.40 & 3.30 & 3.02 & 0.68 & -1.35\\
1.0 & 0.0254 & 4262 & 5403 & 4351 & 2940 & 0.06 & -0.21 & 1.37 & 3.40 & 3.57 & 1.87 & -4.92 & -4.49\\
%1.1 & 0.0045 & TTTT & TTTT & TTTT & TTTT & LLL & LLL & LLL & LLL & AAA & AAA & AAA & AAA\\
1.1 & 0.0143 & 4491 & 5966 & 4455 & 3189 & 0.29 & 0.09 & 1.57 & 3.40 & 3.30 & 3.15 & 1.13 & -0.45\\
1.2 & 0.0143 & 4580 & 6231 & 4469 & 3230 & 0.41 & 0.27 & 1.62 & 3.39 & 3.30 & 3.22 & 1.34 & 0.12\\
1.3 & 0.0143 & 4651 & 6510 & 4474 & 3267 & 0.53 & 0.43 & 1.68 & 3.39 & 3.30 & 3.26 & 1.44 & 0.50\\
1.4 & 0.0143 & 4708 & 6863 & 4477 & 3300 & 0.63 & 0.58 & 1.73 & 3.39 & 3.30 & 3.28 & 1.49 & 0.78\\
1.5 & 0.0143 & 4754 & 7305 & 4486 & 3331 & 0.73 & 0.71 & 1.78 & 3.39 & 3.30 & 3.29 & 1.51 & 0.98\\
1.6 & 0.0143 & 4789 & 7768 & 4493 & 3361 & 0.82 & 0.83 & 1.82 & 3.39 & 3.30 & 3.29 & 1.53 & 1.13\\
1.8 & 0.0143 & 4838 & 8628 & 4462 & 3433 & 0.99 & 1.05 & 1.97 & 3.36 & 3.30 & 3.30 & 1.53 & 1.32\\
2.0 & 0.0143 & 4866 & 9388 & 4494 & 3570 & 1.14 & 1.23 & 2.01 & 3.24 & 3.30 & 3.30 & 1.51 & 1.40\\
2.2 & 0.0143 & 4878 & 10091 & 4373 & 3886 & 1.28 & 1.39 & 2.26 & 2.89 & 3.30 & 3.30 & 1.50 & 1.46\\
\hline
\end{tabular}
\end{table*}

\begin{table*}
\caption{Input parameters for the best fit models along with lithium abundances at the start of the different evolutionary phases for the three ({\it M}, [Fe/H]) cases studied in details in Section \ref{sec:ConstThermoMixing}.\label{table:2}}
\footnotesize
\begin{tabular}{cc|ccc|c|cccc}
    \hline
Mass & [Fe/H] & \multicolumn{3}{c|}{Model parameters} & $\delta A$(Li) & \multicolumn{4}{c}{{\it A}(Li) after correction for $\delta A$(Li) at}\\
 (M$_\odot$) & (dex) & Z & $\alpha_{\rm ML}$ & $\alpha_{\rm th }$ & (dex) & PMS & ZAMS & LB & RGB-tip\\
\hline
1.1 & -0.50 & 0.0045 & 1.80 $\pm$ 0.05 & 150 $\pm$ 50 & -0.3 & 2.48 & 2.48 & 0.92 & -0.77 $\pm$ 0.3\\
1.1 & 0.00 & 0.0143 & 1.85 $\pm$ 0.05 & 100 $\pm$ 50 & -0.6 & 2.70 & 2.55 & 0.53 & -1.05 $\pm$ 0.3\\
1.5 & -0.25 & 0.0080 & 1.95 $\pm$ 0.05 & 300 $\pm$ 100 & -0.2 & 2.84 & 2.83 & 1.10 & -0.08 $\pm$ 0.3\\
\hline
\end{tabular}
\end{table*}

%%%%%%%%%%%%%%%%%%%%%%%%%%%%%%%%%%%%%%%%%%%%%%%%%%

% Don't change these lines
\bsp	% typesetting comment
\label{lastpage}
\end{document}